\documentclass[12pt]{article}
\usepackage[dvips]{graphicx}
\usepackage{amssymb}
\usepackage{amsmath}
\usepackage{epsfig}
\usepackage{cite}
\usepackage{hyperref}
\usepackage{bbold}
\usepackage{multirow}

\numberwithin{equation}{section}
\numberwithin{table}{section}

\def\beq{\begin{equation}}
\def\eeq{\end{equation}}
\def\be{\begin{equation}}
\def\ee{\end{equation}}
\def\bea{\begin{eqnarray}}
\def\eea{\end{eqnarray}}

\def\g{{\mathfrak{g}}}

\DeclareRobustCommand{\SkipTocEntry}[4]{}
\RequirePackage{color}
\newcommand{\cT}{\mathcal{T}}

\newcommand{\cL}{\mathcal{L}}

\newcommand{\cM}{\mathcal{M}}
\newcommand{\cN}{\mathcal{N}}

\newcommand{\cH}{\mathcal{H}}

\newcommand{\cF}{\mathcal{F}}

\newcommand{\cV}{\mathcal{V}}
\newcommand{\cQ}{\mathcal{Q}}

\newcommand{\D}{\mathrm{D}}

\newcommand{\id}{{\mathbb1}}    
\def\AdS{\textrm{AdS}}

\begin{document}

\begin{titlepage}
\begin{center}
\rightline{\small ZMP-HH/16-15}

\vskip 1cm

{\Large \bf Classification of maximally supersymmetric backgrounds in supergravity theories}
\vskip 1.2cm

{\bf  Jan Louis %$^{a,b}$ 
and Severin L\"ust }

\vskip 0.8cm

{\em Fachbereich Physik der Universit\"at Hamburg, Luruper Chaussee 149, 22761 Hamburg, Germany}
\vskip 0.3cm

{\em Zentrum f\"ur Mathematische Physik,
Universit\"at Hamburg,\\
Bundesstrasse 55, D-20146 Hamburg, Germany}
\vskip 0.3cm

\vskip 0.3cm

{\tt jan.louis@desy.de, severin.luest@desy.de}

\end{center}

\vskip 1cm

\begin{center} {\bf ABSTRACT } \end{center}

\noindent

We study maximally supersymmetric solutions of all gauged or deformed 
supergravity theories in $D \ge 3$ space-time dimensions.
%, including gauged supergravity theories.
%Hereby we try to adopt a generic language and aim to be as much independent from details depending on the number of dimensions and supersymmetries.
For vanishing background fluxes the space-time  background has to be 
either Minkowski or anti-de Sitter.
We derive a  simple criterion for the existence of
solutions with non-trivial fluxes 
and determine all supergravities that satisfy it.
We show that their solutions coincide with those
of the corresponding ungauged theories and conclude
that the known list of maximally supersymmetric solutions
is exhaustive.

\vfill

%\today

July 2016

\end{titlepage}

%%%%%%%%%%%%%%%%%%%%%%%%%%%%%%%%%%%%%%%%%%%%%%%%%%%%

%\tableofcontents

%%%%%%%%%%%%%%%%%%%%%%%%%%%%%%%%%%%%%%%%%%%%%%%%%%%%

%%%%%%%%%%%%%%%%%%%%%%%%%%%%%%%%%%%%%%%%%%%%%%%%%%%%
\section{Introduction}

Classical solutions of supergravity theories have been intensely studied. 
%and in some cases
%, especially AdS solutions received lots of attention since the advent of the AdS/CFT correspondence \cite{Maldacena:1997re}. \Snote{AdS/CFT erwaehnen?}
%many solutions %for different theories in various dimensions 
%with different amounts of supersymmetry
%have been 
%classified.
%(For an overview see, for example,
 % \cite{AlonsoAlberca:2002dw,Gauntlett:2005bn} and references therein.)
%\cite{Tod:1983pm, Gauntlett:2002nw, FigueroaO'Farrill:2002ft, Gauntlett:2003fk, Gutowski:2003rg, Chamseddine:2003yy, Caldarelli:2003pb}.
Of particular interest are supersymmetric bosonic solutions where
the supersymmetry variations of all fermionic fields present in the theory vanish.
These have been classified in many cases, nevertheless a complete description of all supersymmetric solutions has not been obtained yet.%
\footnote{For the ungauged case see for example
\cite{Tod:1983pm, Behrndt:1997ny, Gauntlett:2002sc, Gauntlett:2002nw, FigueroaO'Farrill:2002ft, Gauntlett:2002fz, 
Gauntlett:2003cy, Gutowski:2003rg, Chamseddine:2003yy, Gauntlett:2003wb, Gauntlett:2004zh, Dall'Agata:2004dk, 
Cariglia:2004qi, Gauntlett:2004qy,Gillard:2004xq, Gauntlett:2004hs, Gran:2005wn, Gutowski:2005id,
Bellorin:2005zc, Ishino:2005ru, Meessen:2006tu, Huebscher:2006mr, Bellorin:2006yr,
Cacciatori:2007vn, Meessen:2010fh, Akyol:2010iz} and
\cite{Gauntlett:2003fk, Caldarelli:2003pb, Gutowski:2004yv, Cariglia:2004kk, Cacciatori:2004rt, Bellorin:2007yp,
Cacciatori:2008ek, Hubscher:2008yz, Bellorin:2008we, Klemm:2009uw, Hristov:2009uj, Klemm:2010mc, Meessen:2012sr} for the gauged case.}
In this paper we exclusively focus on maximally supersymmetric solutions, 
that is solutions 
which preserve all supercharges of a given supergravity.
In this case 
the resulting Killing spinor equations have to admit 
an independent Killing spinor for each supercharge which
considerably  constrains the allowed space-time backgrounds.

If one excludes background fluxes, 
the Killing spinor equations take a particularly simple form 
and can be integrated directly. In this case one finds that in
supergravities with $D$ space-time dimensions
only two backgrounds are possible: 
$D$-dimensional Minkoswki space~\(M_D\)  
or $D$-dimensional anti-de Sitter space \(\AdS_D\).
While \(M_D\) is a solution of all ungauged supergravities, \(\AdS_D\)
backgrounds require a non-trivial scalar potential and hence the supergravities
have to be gauged or otherwise deformed.\footnote{By deformed supergravities we denote
  theories
which are  deformed by mass parameters or the
superpotential in $D=4,\, \cN=1$ supergravity, for example.}
It turns out that generically additional algebraic conditions have to be satisfied
which further restrict the gauged and/or deformed
supergravity.
%(see, for example, \cite{deAlwis:2013jaa,Louis:2014gxa,Louis:2015mka,Louis:2015dca,Louis:2016qca} \Jnote{complete list?} and references therein).

More complicated solutions only arise if non-trivial background fluxes of 
gauge potentials in the gravitational multiplet are turned on.
However, these fluxes generically enter the supersymmetry variation of 
the spin-1/2 fermions in the gravitational multiplet and break supersymmetry 
(at least partially). The only exception occurs for gauge potentials 
with (anti-)self-dual field strengths in a chiral theory which drop out of
the spin-1/2 variations.
%We therefore conclude that background fluxes which do not break any supercharge
%are only possible in a very limited set of supergravities.
As we will see, this limits the  possible supergravities to a small subset where
either the gravitational multiplet does not contain spin-1/2 fermions or the theory is chiral and features (anti-)self-dual fields.
%We list all supergravities satisfying this requirement in table~\ref{tab:pformfluxes}. 

We further show that in the presence of non-trivial fluxes 
the background values of the  supersymmetry variations 
have to coincide with the  supersymmetry variations of the
corresponding ungauged theories.
Concretely this means that the fermionic shift matrices as well as the
gauge connection for the R-symmetry have to vanish in the background.\footnote{This result has been obtained previously 
for gauged \(D=4\), \(\cN=2\) supergravity in \cite{Caldarelli:2003pb, Hristov:2009uj}, for minimal gauged \(D=5\), \(\cN=2\) supergravity in \cite{Gauntlett:2003fk} and for \(D=6\), \(\cN = (1,0)\) supergravity in \cite{Akyol:2010iz}.
However theses results always rely on the specific formulation of the
particular gauged supergravities under consideration.}
We prove this property by deriving a generic expression for the
R-symmetry connection. As a by-product we show that in maximally
supersymmetric backgrounds the R-symmetry can only be gauged by vector
fields of the gravitational multiplet.%
\footnote{In the generic situation it is precisely speaking not the R-symmetry which is being gauged but a subgroup of the scalar field space's isometry group which in turn induces R-symmetry transformations.
For a more detailed discussion see the explanations around \eqref{eq:rgenerators} and also in appendix~\ref{app:rconnection}.}

The correspondence with the ungauged theories implies that also the background
solutions coincide. It turns out that for all situation where
background fluxes are possible  the maximally supersymmetric solutions  have already been determined and classified \cite{Tod:1983pm, Gauntlett:2002nw, FigueroaO'Farrill:2002ft, Gutowski:2003rg, Chamseddine:2003yy}.
They are either 
space-times of the  Freund-Rubin form \(AdS_{d} \times S^{(D-d)}\) \cite{Freund:1980xh} 
or H{\it pp}-wave solutions \cite{KowalskiGlikman:1984wv,KowalskiGlikman:1985im}.
Only in five space-time dimensions one can have more exotic
solutions~\cite{Gauntlett:2002nw}.  
Although all these solution were known previously our analysis shows
that this list is exhaustive.

This paper is organized as follows.
In section~\ref{sec:prelim} we set the stage for our analysis and recall
the supersymmetry transformations of the fermionic fields.
In particular we establish a notation which allows us to discuss
the various supergravities in a common framework.
In section~\ref{sec:susybackgrounds} we show that supergravities 
with \(D\)-dimensional space-times without fluxes have supersymmetric backgrounds that are either $\AdS_D$ or $M_D$.
%In particular $\AdS_d$ backgrounds for $d<D$ do not exist in this case. \Snote{Brauchen wir diesen Satz noch?}
In section~\ref{sec:susyflux} we turn on background fluxes 
and argue that the vanishing of the supersymmetry transformation of the 
spin-$\tfrac12$ fermions
requires that all background fluxes are zero except in chiral theories with 
(anti-) self-dual fluxes.
%Therefore backgrounds other  than \(AdS_D\) and \(M_D\) can only exist in a few exceptional situations where either the gravitational multiplet does not contain spin-$\frac12$ fermions or the background flux is (anti-)self-dual.
We further show that solutions with non-trivial fluxes 
coincide with the solutions of the corresponding ungauged theories.
%We recall the existing results from
%In section~\ref{sec:AdSresult} we show that 
%preserving all supercharges of a given $D$-dimensional supergravity constrains 
%$Y^{(D-d)}$ has to be the sphere $S^{(D-d)}$
%and the non-vanishing background flux has to be proportional to the top-form 
%on at least one of the two factors.
%This reduces the possible supersymmetric AdS-backgrounds to a finite list.
Some of the technical analysis is relegated to three appendices.
In appendix~\ref{app:gamma} we summarize our $\Gamma$-matrix conventions,
in appendix~\ref{app:integrability} we supply some of the technical details
necessary in section~\ref{sec:susyflux} 
and finally in appendix~\ref{app:rconnection} we determine the general gauging of the R-symmetry in maximally supersymmetric backgrounds.

%%%%%%%%%%%%%%%%%%%%%%%%%%%%%%%%%%%%%%%%%%%%%%%%%%%%%%%%%%%%%%%%%%%%%
\section{Preliminaries}\label{sec:prelim}
In this paper we discuss properties of supergravity backgrounds
in arbitrary space-time dimensions and for varying number of supercharges in a given dimension. In order to avoid  a case-by-case analysis
we introduce a unifying notation which allows us to 
more or less discuss all cases simultaneously.
It is the purpose of this section to set the stage for this analysis and 
provide a common notation.

Supergravities in \(D\) space-time dimensions contain a
gravitational multiplet whose generic field content includes  the metric \(g_{MN}\), \(M, N = 0,
\dots, D-1\), \(\cN\) gravitini \(\psi^i_M\), \(i = 1,\dots,\cN\),
a set of \((p-1)\)-form gauge potentials \(A^{(p-1)}\) (with $p$-form field strengths
\(F^{(p)}\)), a set of spin-$\frac12$ fermions $\chi^a$ as well as a set of
scalar fields $\phi$. Note that not all of these component fields
necessarily have to be
part of a given gravitational multiplet but we 
gave the most general situation.
Moreover, there might be additional multiplets in the spectrum
(e.g.\ vector, tensor or matter multiplets) which 
can also have gauge potentials among their component members.
For the moment we denote all gauge potentials by \(A^{(p-1)}\) 
but we distinguish them shortly.
The spin-$\frac12$
fermions in the extra multiplets we collectively call \(\lambda^s\)
while all scalars we universally denote as $\phi$.

Let us first focus on the kinetic terms for the gauge potentials.
They 
%all the form fields from all possible multiplets in the theory 
are of the generic form (for a review see, for example, \cite{Samtleben:2008pe})
\begin{equation}
\cL_{kin} = - \frac{1}{2} \sum_p M^{(p)}_{I_{p} J_{p}}\!\left(\phi\right)\, F^{(p) I_{p}} \wedge \ast F^{(p) J_{p}} \,,
\end{equation}
where the indices \(I_{p}, J_{p}\) label field strengths
of the same rank \(p\)
and the sum runs over all possible \(p\)-forms 
which are present in a given theory.
The matrices \(M^{(p)}\) generically  depend on all scalar fields, are symmetric and positive definite.
Therefore they can be diagonalized via
\begin{equation}
M^{(p)}_{I_pJ_p} = \delta_{\alpha_p\beta_p} \cV^{\alpha_p}_{I_p}  \cV^{\beta_p}_{J_p}  \,,
\end{equation}
where the vielbeins \(\cV^{\alpha_p}_{I_p}\) are again scalar dependent.
Later on it will be important to distinguish which of the form fields enter the supersymmetry variations of the gravitini.
For this purpose we introduce the abbreviation\footnote{In the following 
we frequently drop the labels \((p)\) in order to not overload the notation.}
\begin{equation}
F^{\alpha_p} = \cV^{\alpha_p}_{I_p} F^{I_{p}} \, ,
\end{equation}
and split the indices \(\alpha_p\) according to
\begin{equation}\label{indexsplit}
\alpha_p = \left(\hat \alpha_p, \tilde \alpha_p\right) \, .
\end{equation}
We then denote by \(F^{\hat\alpha_p}\) 
the field strengths in the gravitational multiplets 
(e.g.\ the graviphotons for \(p = 2\)) and by  \(F^{\tilde\alpha_p}\) 
the field strengths of gauge potentials which arise in 
all other multiplets that might be present.
Note that this split depends on the scalar fields via the vielbeins $\cV$
and thus is background dependent.

After this preparation we recall the supersymmetry variations of the fermions, which are of special importance in the following.
The transformation of the gravitini takes the generic form
\begin{equation}\label{eq:gravitinovariation}
\delta \psi^i_M = \D_M  \epsilon^i  + \left(\cF_M\right)^i_j \epsilon^j + A_{0\,j}^i \Gamma_M\epsilon^j \ ,
\end{equation} 
where 
\begin{equation}\label{eq:covariantderiv}
\D_M \epsilon^i = \nabla\!_M \epsilon^i - \left(\cQ_M\right)^i_j \epsilon^j \,.
\end{equation}
\(\nabla\!_M\) denotes the Levi-Cevita connection and \(\cQ_M\)
is the %an additional 
R-symmetry connection which we discuss in more detail shortly.
The second term in \eqref{eq:gravitinovariation} contains the various field strengths and is given by
\begin{equation}\label{cFdef}
\big(\cF_M\big)^i_j  =  \sum_{p \geq 2} \sum_{\hat\alpha_p}\big(B^{(p)}_{ \hat\alpha_p}\big)^i_j\,
%F^{(p)} \cdot T_M = \alpha_{(p)} 
F^{\hat\alpha_p}_{{N_1}\dots {N_p}} {T_{(p)}^{{N_1}\dots {N_p}}}{}_M \,,
\end{equation}
where the \(B^{(p)}\) are constant  matrices. %and will be determined in appendix~\ref{app:sufficient}. 
%where the sum runs over all form-fields in the gravity multiplet.\footnote{discuss mixing}
%Since in the theories we consider the gravity multiplet does not contain any scalars, the prefactor \(\alpha_{(p)}\) is just a numerical constant, depending on the dimension \(D\) and the value of \(p\).
The matrices \({T_{(p)}^{{N_1}\dots {N_p}}}{}_M\) are defined as
%combination of antisymmetrized products of $\Gamma$-matrices 
\begin{equation}\label{eq:T}
{T_{(p)}^{{N_1}\dots {N_p}}}{}_M = {\Gamma^{{N_1}\dots {N_p}}}{}_M + \beta_{(p)}\, \Gamma^{[N_1\dots N_{p-1}} \delta^{N_p]}_M \,,
\end{equation}
where $\Gamma^{{N_1}\dots {N_p}}$ is an antisymmetrized product of $\Gamma$-matrices (see appendix~\ref{app:gamma} for our conventions) 
and
\begin{equation}\label{eq:beta}
\beta_{(p)} = \frac{p(D-p-1)}{p-1} \,. %\ ,
\end{equation}
%which we also derive in appendix~\ref{app:sufficient}.
Finally,  the matrix \(A_0\) in the third term of 
\eqref{eq:gravitinovariation} arises in gauged and/or deformed supergravities 
and  is parameterized by the gaugings or deformations and in general
depends on all scalar fields in the spectrum \cite{Samtleben:2008pe}.
Its precise form is specific to the supergravity under consideration.

Let us now turn to the spin-$\frac12$ fermions \(\chi^a\) and $\lambda^s$.
The \(\chi^a\) are part of the gravitational multiplet  
and their transformations take the generic form
\begin{equation}\label{eq:spin12variation}
\delta \chi^a = \sum_{p \geq 1} \sum_{\hat\alpha_p}\big(C^{(p)}_{ \hat\alpha_p}\big)^{a}_i \, F^{\hat\alpha_p}_{N_1\dots N_p} \Gamma^{N_1\dots N_p} \epsilon^i + A^a_{1\,i}\, \epsilon^i \ .
\end{equation}
%where again \(\beta_{(p)}\) and \(A_1\) are scalar field dependent matrices.
%\Jnote{zweimal $\beta$}
The $\lambda^s$ are members of other multiplets  present
(e.g.\ vector-, tensor- or matter-multiplets) and we similarly have
\begin{equation}\label{eq:spin12variationb}
\delta \lambda^s = \sum_{p\geq1} \sum_{\tilde\alpha_p}\big(D^{(p)}_{ \tilde\alpha_p}\big)^{s}_i \, F^{\tilde\alpha_p}_{N_1\dots N_p} \Gamma^{N_1\dots N_p} \epsilon^i + A^s_{2\,i}\, \epsilon^i \,.
\end{equation}
Note that the field strengths appearing in \eqref{eq:spin12variation} and \eqref{eq:spin12variationb} form a disjoint set. 
Accordingly they are labeled by \(\hat \alpha_p\),  \(\tilde \alpha_p\) that we introduced in \eqref{indexsplit}.
Contrary to \eqref{cFdef} the sums in \eqref{eq:spin12variation} and
\eqref{eq:spin12variationb} start already at \(p = 1\)
and thus include  the fields strengths   of the scalar fields
\(F^{\alpha_1}_M=D_M\phi^{\alpha_1}\) which
do not enter the gravitino variations \eqref{eq:gravitinovariation}.
As in the gravitino variations \(C^{p}\) and \(D^{p}\) are constant matrices while \(A_1\) and \(A_2\) arise in gauged supergravities
and depend on the gaugings/deformations and the scalar fields. 
They have a specific form in a given supergravity.
%where here the sum runs only over those $p$-form fields which are
%in the same multiplet with $\lambda^s$ and we denote their field strengths
%by \(H^{(p)}\).
%While the matrices \(\alpha_{(p)}\), \(\beta_{(p)}\) and \(\gamma_{(p)}\)  only depend on scalars from the same multiplet as the respective fermions, the  fermionic shift matrices \(A_0\), \(A_1\) and \(A_2\) can generically depend on all scalars of the theory.
Supersymmetry relates the fermionic shift matrices \(A_0\), \(A_1\) and \(A_2\) to the scalar potential \(V\)
and generically one has
\begin{equation}\label{eq:potential}
V = - c_0 \mathrm{tr}(A_0^\dagger A_0) + c_1 \mathrm{tr}(A_1^\dagger A_1) + c_2 \mathrm{tr}(A_2^\dagger A_2) \,,
\end{equation}
where \(c_0\), \(c_1\) and \(c_2\) are numerical constants fixed by
supersymmetry in a given supergravity.

Let us return to the
%R-symmetry
connection \(\cQ_M\) 
in the covariant derivative
\eqref{eq:covariantderiv} as it will play an important role in the following
and we need to establish some of its properties.
In a generic %, possibly gauged 
supergravity \(\cQ_M\)  splits according to \cite{Samtleben:2008pe,Bandos:2016smv,Trigiante:2016mnt}
\begin{equation}\label{eq:qsplit}
\cQ_M = \cQ^\mathrm{scalar}_M + \cQ^\mathrm{gauge}_M \,,
\end{equation}
where \(\cQ^\mathrm{scalar}_M\) is a composite connection which only
depends on the scalar
fields and their derivatives and 
%which
already exists in the ungauged theory.
%In the presence of scalar fields this term is needed since transformations of the scalar fields induce due to supersymmetry also transformations of the fermionic fields.
A transformation along a possible isometry of the scalar field space can induce a scalar field dependent R-symmetry transformation and the connection term \(Q_M^\mathrm{scalar}\) is necessary to make \(\D_M \epsilon^i\) transform covariantly.%
\footnote{To be more specific let us assume that the scalar fields \(\phi\) span some manifold \(\cT\), i.e. \(\phi\colon\cM_D \rightarrow \cT\), where \(\cM_D\) denotes the space-time manifold.
Generically the gravitini and hence also the supersymmetry parameters \(\epsilon^i\) are sections of a (non-trivial) vector bundle over \(\cT\) with connection \(\omega\).
Then \(\cQ^\mathrm{scalar}\) is just the pullback of this connection with respect to \(\phi\), i.e. \(\cQ^\mathrm{scalar} = \phi^{\ast} \omega\).}
These transformations can be made local (i.e. not only scalar field but also explicitly space-time dependent) by introducing another term
\(\cQ^\mathrm{gauge}_M\) which contains a linear combination of gauge
fields \(A^{\alpha_2}\), i.e. 
\begin{equation}\label{eq:rgenerators}
\cQ^\mathrm{gauge}_M = A^{\alpha_2}_M t_{\alpha_2} \,.
\end{equation}
The matrices \(t_{\alpha_2}\) are often called moment maps and generically depend again on the scalar fields. 
They take values in the Lie-algebra \(\g_R\) of the R-symmetry group, but at most points in field space they do not need to span a proper Lie-subalgebra of \(\mathfrak{g}_R\).
However, we will show in appendix~\ref{app:rconnection} that the \(t_{\alpha_2}\) close under the action of the Lie-bracket in every maximally supersymmetric background.
%where the generators \(t_{\alpha_2}\) span the gauged subgroup of the
%R-symmetry group and, as we will see, depend on the scalar fields.
In this specific situation we
% will 
%from now on 
say for simplicity
that the subgroup of the R-symmetry group which is generated by the background values of the \(t_{\alpha_2}\) is gauged \emph{in this background}, or shortly that the R-symmetry is gauged (in this background).
As explained above it would be however generically more precise to speak about gauging a certain subgroup of the scalar manifold's isometry group which in turn induces R-symmetry transformations.
Moreover, let us 
stress that in principle all gauge fields, those from the gravitational multiplet
(the graviphotons) as well as gauge fields from other multiplets
(e.g.\  vector multiplets),
can appear in \eqref{eq:rgenerators}.
%In summary one could also think of \(Q_M\) as the ``covariantization'' of \(Q^\mathrm{scalar}_M\) with respect to the gauging described by \eqref{eq:rgenerators}.

In the following we also need the curvature or field strength
\(\cH_{MN}\) of \(\cQ_M\). As usual it appears in the
commutator
of the covariant derivatives defined in \eqref{eq:covariantderiv} as follows
\begin{equation}\label{eq:Dcomm}
\left[\D_M, \D_N \right] \epsilon^i = \tfrac{1}{4} R_{MNPQ}\Gamma^{PQ}\, \epsilon^i - \left(\cH_{MN}\right)^i_j \epsilon^j \,,
\end{equation}
where \(R_{MNPQ}\) is the Riemann tensor of the background space-time $\cM_D$.
As a consequence of \eqref{eq:qsplit}  \(\cH_{MN}\) similarly decomposes as
\begin{equation}\label{eq:rfieldstrength}
\cH_{MN} = \cH^\mathrm{scalar}_{MN} + \cH^\mathrm{gauge}_{MN} \, ,
\end{equation}
with %\(\cH^\mathrm{gauge}_{MN}\) %is a linear combination of the field strengths \(F^{\alpha_2}_{MN}\) of the gauge fields,
\begin{equation}\label{eq:hrgenerators}
\cH^\mathrm{gauge}_{MN} = F^{\alpha_2}_{MN} t_{\alpha_2} \,,
\end{equation}
and  \(t_{\alpha_2}\) being the same matrices as in
\eqref{eq:rgenerators}. 
The field strength of the composite connection
\(\cH^\mathrm{scalar}_{MN}\) can be expressed in terms of the field
strengths of the scalar fields  \(F^{\alpha_1}_M\)
and takes the generic form 
\begin{equation}\label{eq:hscalar}
\cH^\mathrm{scalar}_{MN} = h_1 C^\dagger_{\hat\alpha_1} C_{\hat\beta_1} F^{\hat\alpha_1}_{[M} F^{\hat\beta_1}_{N]} + h_2 D^\dagger_{\tilde\alpha_1} D_{\tilde\beta_1} F^{\tilde\alpha_1}_{[M} F^{\tilde\beta_1}_{N]} \,,
\end{equation}
where $C,D$ are the matrices appearing in \eqref{eq:spin12variation}
for $p=1$ and
\eqref{eq:spin12variationb} respectively and \(h_1\) and \(h_2\) are
numerical constants determined by supersymmetry in a given supergravity.

Let us close this section by recalling that a 
supergravity background which preserves some 
supersymmetry has to admit spinors \(\epsilon^i\) which
satisfy
\begin{equation}
\delta \psi^i_M = \delta \chi^a = \delta \lambda^s = 0 \ .
\end{equation}
The number of linearly independent such spinors then
determines the number of preserved supercharges.
In this paper we only consider backgrounds which preserve all
supercharges of the supergravity under consideration. 
This considerably  simplifies the analysis as we will see shortly.

\section{Supersymmetric backgrounds without fluxes}\label{sec:susybackgrounds}

Let us first analyze the situation where all background fluxes vanish and hence
eqs.~\eqref{eq:gravitinovariation}--\eqref{eq:spin12variationb}
simplify. 
If all supercharges are preserved, 
\(\delta \chi^a = \delta \lambda^s = 0\) imply 
via \eqref{eq:spin12variation} and \eqref{eq:spin12variationb} 
that\footnote{Both equations only have to hold in the background,
i.e.\ the conditions read $\langle {A_1}\rangle=\langle {A_2}\rangle=0$.
However, in order to keep the notation manageable 
we generically omit the brackets henceforth.}
\begin{equation}
A_1 = A_2 = 0\ .
\end{equation}
On the other hand, the vanishing of the gravitino variation \eqref{eq:gravitinovariation}
\begin{equation}\label{eq:susyvariation}
\delta \psi^i_M = \D_M \epsilon^i + A_{0\,j}^i \Gamma_M \epsilon^j = 0
\end{equation}
says that $\epsilon^i$ has to be a Killing spinor.
Its existence implies a strong constraint on the space-time manifold
which can be derived by 
acting with another covariant derivative, antisymmetrizing
and using \eqref{eq:Dcomm}. 
%\begin{equation}
%\nabla\!_N \nabla\!_M \epsilon^i - A^i_{0\,j} A^j_{0\,k} \Gamma_M \Gamma_N \epsilon^k = 0 \,.
%\end{equation}
This  implies
\begin{equation}\label{Riemann}
\biggl[\left(\tfrac{1}{4} {R_{MN}}^{PQ} \delta^i_k+ 2 A^i_{0\,j} A^j_{0\,k} \delta^P_M \delta^Q_N \right) \Gamma_{PQ} + 2\left(\D_{[M} A_0 \right)^i_k\Gamma_{N]} \biggr]\epsilon^k = 0 \ ,
\end{equation} 
where we also used that $\cH_{MN}$ vanishes in backgrounds without any fluxes
and where 
the covariant derivative of \(A_0\) is defined as
  \(\D_M A_0 = \partial_{M} A_0 - \left[\cQ_{M}, A_0\right]\).
In a background which preserves all supercharges the expression in the
bracket has to vanish at each order in the \(\Gamma\)-matrices independently.
From the term linear in \(\Gamma\) we learn that \(A_0\) is covariantly
constant.
The part quadratic in \(\Gamma\) then says that \(A^2_0\) needs to be proportional to the identity matrix and must be a constant since
\begin{equation}
\partial_M A_0^2 = \D_M A_0^2 = 0 \,.
\end{equation}
Moreover it implies that in a given supergravity the maximally supersymmetric backgrounds
have to be maximally symmetric space-times with
a Riemann tensor given by
\begin{equation}\label{eq:riemann}
R_{MNPQ} = - \frac{4}{\cN}\ \mathrm{tr}\!\left(A_0^2\right) \left(g_{MP} g_{NQ} - g_{MQ}g_{NP}\right) \,.
\end{equation}
From the canonical Einstein equations one readily infers that in such backgrounds the cosmological constant \(\Lambda\) is given by
\begin{equation}\label{LambdaA}
\Lambda = - \frac{2}{\cN}\, (D-1)(D-2)\, \mathrm{tr}\!\left(A^2_0\right) \,,
\end{equation}
and the background value of the 
scalar potential is related by \(\left<V\right> = \Lambda\).
Note that consistency then determines the coefficient $c_0$ of $V$ in \eqref{eq:potential} to be 
$c_0 = \frac{2}{\cN}\,(D-1)(D-2)$.
For $A_0\neq0$ we thus have an  AdS-background $\cM_D=\AdS_D$ while for 
$A_0=0$ the background is flat. 
So altogether fully supersymmetric backgrounds without background fluxes have to be one of the following cases
\begin{equation}\label{eq:firstresult}
 \cM_D=\AdS_D   \qquad \textrm{or} \qquad  
\cM_D = M_d \times T^{(D-d)} \ ,\qquad 1 \leq d \leq D 
%\ ,\qquad \cM_D = M_D %\times T^{(D-d)} \ .
\ ,
\end{equation} 
up to local isometries.
%where \(M_d\) is a Minkowskian manifold, $T^{(D-d)}$  is a torus and . \Snote{Geht z.B. \(d=0\)?} 
We see in particular that without fluxes supersymmetric backgrounds with an $\AdS_{d}$ factor cannot exist for $d<D$.

Before we proceed let us note that in a given $D$-dimensional 
gauged supergravity
the existence of the $\cM_D=\AdS_D$ background requires the existence 
of a solution with
\begin{equation}\label{AdSsolution}
A_0^2= -\tfrac{\Lambda}{2(D-1)(D-2)}\, \mathbb{1}\ ,\qquad A_1 = A_2 = 0\ .
\end{equation}
This can only be checked in a case-by-case analysis and explicit
solutions have indeed been constructed in a variety of supergravities
(see, for example, \cite{Hristov:2009uj,deAlwis:2013jaa,Louis:2014gxa,Louis:2015mka,Louis:2015dca,Louis:2016qca} and references therein). However, from ref.~\cite{Nahm:1977tg} it is known that AdS superalgebras only exist for 
$D<8$ and in $D=6$ only for the non-chiral $\cN = (1,1)$ supergravity. 
In the other cases no solution of \eqref{AdSsolution} can exist.

%%%%%%%%%%%%%%%%%%%%%%%%%%%%%%%%%%%%%%%%%%%%%%%%%%%%%%%%%%%%%%%%%%%%%%%%%
\section{Supersymmetric backgrounds with fluxes}\label{sec:susyflux}

In this section we extend our previous analysis in that we
consider backgrounds with non-trivial
fluxes and reanalyze the implications for the possible
space-time manifolds.
%In the previous section we observed that 
%As we have seen above, solutions different from \(\mathrm{Mink}_d \times T^{(D-d)}\), with \(0 \leq d \leq D\), or \(AdS_D\) can only be obtained if \eqref{eq:susyvariation} is modified by the introduction of non-vanishing background fluxes.
In this case  the vanishing supersymmetry variations of the spin-1/2
fermions  given in
\eqref{eq:spin12variation} and \eqref{eq:spin12variationb} 
immediately impose additional constraints.
As we will see, they are particularly strong for the fermions 
$\chi^a$ in the gravitational multiplet.
% give rise to constraints on theories where this is possible without breaking supersymmetry.
Since the $\Gamma$- matrices and their antisymmetric products are
linearly independent, \(\delta \chi^a=\delta \lambda^s=0\) enforces
\begin{equation}\label{eq:spin12conditions}
A_1 = A_2 = 0 \qquad \textrm{and}\qquad F^{(p)} = 0\ , 
\end{equation}
for all possible values of \(p\).\footnote{In even dimensions \(D\)
  all antisymmetric products of gamma matrices are linearly
  independent while in odd dimensions only those up to rank
  \((D-1)/2\) are linearly independent as can bee seen from
  \eqref{eq:gammahodgeodd}. This however is strong enough to enforce \eqref{eq:spin12conditions}.}
This seems to imply that no background fluxes can be turned on.
However, this conclusion  can be evaded either if  there simply are
no  spin-1/2 fermions in the gravity multiplet or 
if there is an (anti-) self-dual field strength in a chiral theory. 

In the first case there is no condition on the fluxes $F^{\hat\alpha_p}$
which appear in the gravitino variation \eqref{eq:gravitinovariation} and
\eqref{cFdef} but only on the fluxes $F^{\tilde\alpha_p}$ which feature
in \eqref{eq:spin12variationb}.
The second exception follows from the definition of the chirality operator  
\(\Gamma_\ast\)
(given in \eqref{eq:gamma5}) which implies that 
in even dimensions \(D\) the Hodge-dual of a \(p\)-form \(F^{(p)}\) satisfies 
\begin{equation}\label{eq:Fdual}
\ast F^{(p)} \cdot \Gamma = -(-1)^{p(p-1)/2} i^{D/2 + 1} \left(F^{(p)} \cdot \Gamma\right) \Gamma_\ast \ ,
\end{equation}
where we abbreviated
\(F^{(p)} \cdot \Gamma=F_{N_1\dots N_p}^{(p)} \Gamma^{N_1\dots N_p}\)
(and used \eqref{eq:gammahodgeeven}).
Note that the prefactor is real in dimensions \(D = 2 \mod 4\), which are precisely those dimensions in which chiral theories can exist.
In these dimensions one finds for an \mbox{(anti-)} self-dual \(D/2\)-form \(F_{\pm} = \pm \ast F_{\pm}\) that
\begin{equation}\label{eq:FGselfdual}
F_{\pm} \cdot \Gamma = \left(F_{\pm} \cdot \Gamma\right) P_{\pm} \,,
%\begin{cases}
%\left(F_{\pm} \cdot \Gamma\right) P_{\mp} & \mbox{for } D = 2 \mod 8 \\
%\left(F_{\pm} \cdot \Gamma\right) P_{\pm} & \mbox{for } D = 6 \mod 8
%\end{cases} \quad ,
\end{equation}
where \(P_\pm = \frac{1}{2}\left(\id \pm \Gamma_\ast\right)\).
In the chiral supergravities in $D=6, 10$ \cite{Schwarz:1983wa,Schwarz:1983qr,
Nishino:1984gk,Awada:1985er}  the supergravity multiplet contains %(among other fields) 
two or four-form fields, respectively, with self-dual field strengths \(F_+^{\hat\alpha_{D/2}}\).
In these theories the gravitini and consequently also the supersymmetry parameters \(\epsilon^i\)
are left-handed. Therefore, a term of the form \((F^{\hat\alpha_{D/2}}_+\cdot \Gamma)\, \epsilon^{i-}\) cannot appear in \eqref{eq:spin12variation} which indeed shows that a non-vanishing background value for a self-dual field strength does not break supersymmetry in these theories.
%More specifically, in six dimensions \((F_+ \cdot \Gamma) \epsilon\) vanishes for \(\epsilon\) being a left-handed spinor, while in ten dimensions \((F_+ \cdot \Gamma) \epsilon\) vanishes if \(\epsilon\) is right-handed.
Nevertheless \(F_+^{\hat\alpha_{D/2}}\) still enters the variation of the gravitini, as a different contraction with \(\Gamma\)-matrices appears in \eqref{cFdef}.
Hence maximally supersymmetric solutions with non-trivial background flux are possible.
 
The previous considerations in this section enable us to conclude
that solutions which preserve all supercharges of a given supergravity
and which are 
different from the ones
described 
in the previous section 
can only exist if at least one of the following two conditions hold:

\emph{Either the gravity multiplet contains $p$-form gauge fields but 
no spin-$\frac12$ fermions~$\chi^a$ 
or the theory is chiral and (some of) 
the gauge potentials in the gravity multiplet 
satisfy an (anti-) self-duality condition such that they drop out of 
$\delta\chi^a$.}

In table~\ref{tab:pformfluxes} we list all possible supergravities in dimensions \(D \geq 3\) which satisfy these conditions, together with the possible background fluxes.%
\footnote{It is in fact easy to see that such theories cannot exist in \(D=3\) dimensions.
Since three-dimensional gravity is non-dynamical, the graviton, and via supersymmetry also the gravitini, do not carry any on-shell degrees of freedom.
So whenever the gravity multiplet contains vector or scalar fields (which are dual in three dimensions) it must also contain spin-1/2 fields as supersymmetric partners.}
%\footnote{Note that spin-$\frac12$ fermions which are part of matter or gauge multiplets do not affect this argument. In their variations \eqref{eq:spin12variationb} only field strengths contribute which reside in the same multiplet  and they similarly have to vanish. In addition they cannot contribute to the AdS-background as they do not appear  in  the gravitino variations \eqref{eq:gravitinovariation}.}
%In the next section we are going to determine the concrete structure of the AdS-backgrounds.
We now proceed by analyzing the supersymmetry variation of the gravitini \eqref{eq:gravitinovariation} for these theories 
%in table~\ref{tab:pformfluxes} 
in more detail.
%In all the theories listed in table~\ref{tab:pformfluxes}

\begin{table}[htb]
\centering
\begin{tabular}{|c|c|c|c|c|}
\hline
dimension & supersymmetry & q %\# real supercharges 
& possible flux & ref. \\
\hline
\(D = 11\) &  \(\cN = 1\) & 32 & \(F^{(4)}\) & \cite{FigueroaO'Farrill:2002ft} \\
\(D = 10\) & IIB & 32 & \(F_+^{(5)}\) & \cite{FigueroaO'Farrill:2002ft} \\
\(D = 6\) & \(\cN = (2,0)\) & 16 & \(5 \times F_+^{(3)}\) & \cite{Chamseddine:2003yy} \\
\(D = 6\) & \(\cN = (1,0)\) & 8 & \(F_+^{(3)}\) & \cite{Gutowski:2003rg} \\
\(D = 5\) & \(\cN = 2\) & 8 & \(F^{(2)}\) & \cite{Gauntlett:2002nw} \\
\(D = 4\) & \(\cN = 2\) & 8 & \(F^{(2)}\) &  \cite{Tod:1983pm} \\
\hline
\end{tabular}
\caption{Supergravity theories which allow for a background flux that does not break supersymmetry. $q$ denotes the number of real supercharges. In the last column we give the reference for the classification of maximally supersymmetric solutions.}
\label{tab:pformfluxes}
\end{table}

Taking a covariant derivative of \eqref{eq:gravitinovariation} and using \eqref{eq:Dcomm} we arrive at the integrability condition
\begin{equation}\begin{split}\label{eq:fluxgaugeintegrability}
\biggl(&\frac{1}{4}R_{MNPQ} \Gamma^{PQ} \delta^i_j - \left(\cH_{MN}\right)^i_j + 2 \left(\D_{[M} \cF_{N]} + \D_{[M} A_0 \Gamma_{N]}\right)^i_j \\
&\qquad\qquad+ \left[\left(\cF_M + A_0 \Gamma_M\right)^i_k\left(\cF_N + A_0 \Gamma_N\right)^k_j - (M \leftrightarrow N)\right]\biggr) \epsilon^j = 0 \,.
\end{split}\end{equation}
In a maximally supersymmetric background this has to vanish at each order in the \(\Gamma\)-matrices independently.
As we show in appendix~\ref{app:integrability} for all the theories in table~\ref{tab:pformfluxes} the only term at zeroth order in $\Gamma$ is \(\cH_{MN}\)
and thus we arrive at
\begin{equation}\label{eq:rfsvac}
\cH_{MN} = 0 \,.
\end{equation}
Furthermore, due to \eqref{eq:spin12conditions} all scalar fields have vanishing field strengths,
\(F^{\hat\alpha_1} = F^{\tilde\alpha_1} = 0\),
and therefore, using  \eqref{eq:hscalar}, \(\cH^\mathrm{scalar}_{MN}\) automatically vanishes. From \eqref{eq:rfieldstrength} we then learn that 
\eqref{eq:rfsvac} implies
\begin{equation}\label{eq:hgaugecondition}
\cH^\mathrm{gauge}_{MN} = 0 \,.
\end{equation}

In a next step we show that \eqref{eq:hgaugecondition} says
that there can be either no background fluxes at all or that
alternatively both \(A_0\) and \(\cQ^\mathrm{gauge}_M\) vanish in the
background.  
%and the supergravity is effectively \Jnote{??} ungauged. 
To see this 
we derive in appendix~\ref{app:rconnection} that the supersymmetry 
conditions \(A_1 = A_2 = 0\) of \eqref{eq:spin12conditions} enforce \(\cH^\mathrm{gauge}_{MN}\) to be of the generic form
\begin{equation}\label{eq:hgauge}
\cH^\mathrm{gauge}_{MN} \sim F^{\hat\alpha_2}_{MN} \bigl\{A_0, B_{\hat\alpha_2}\bigr\} \,,
\end{equation}
where the precise factor of proportionality is given in \eqref{eq:maxsusyrgenerators} but is not important for the following discussion.
Due to \eqref{eq:rgenerators} and \eqref{eq:hrgenerators} the same relation holds for \(Q^\mathrm{gauged}_M\) with \(F^{\hat\alpha_2}_{MN}\)  replaced by \(A^{\hat\alpha_2}_M\).

Eq.~\eqref{eq:hgauge} has a few notable features.
First of all the appearance of  $F^{\hat\alpha_2}_{MN}$
says that in the background the R-symmetry can only be gauged (in the sense discussed below \eqref{eq:rgenerators}) by graviphotons, i.e.\ by vector fields in the gravity multiplet.%
%Conversely, gauging the R-symmetry
%by vector fields from other multiplets necessarily breaks 
%supersymmetry.
\footnote{This is intuitively plausible but we are not aware
of any previous general proof. Moreover notice that this is a priori only a statement about the maximally supersymmetric background.
At an arbitrary point in field space other gaugings might be in principle possible. See also the discussion at the beginning of appendix~\ref{app:rconnection}.
} 
%non-trivial statement.
Moreover, \eqref{eq:hgauge} uniquely determines the gauged subgroup of the R-symmetry group for maximally supersymmetric vacua and gives an explicit formula for its computation in terms of \(A_0\) and \(B_{\hat\alpha_2}\).
We finally want to stress that this is a generic result, not restricted to the theories in table~\ref{tab:pformfluxes} but true for all gauged supergravity theories with $D\ge 4$. 
%It is an interesting  exercise to check the validity of this statement explicitely for the various gauged supergravity theories known from the literature.

Let us study the implications of \eqref{eq:hgauge} for the supergravities 
of table~\ref{tab:pformfluxes}. We already showed that
the theories which are not in this list cannot have non-vanishing background fluxes so that \eqref{eq:hgaugecondition} is trivially satisfied and does not 
impose any conditions on \(A_0\). Similarly, for
the first three theories in the table~\ref{tab:pformfluxes} it is known that 
deformations by a non-vanishing \(A_0\) do not exist.
In addition no massless vector fields appear 
in the gravitational or in any other multiplet.
Hence \(H^\mathrm{gauge}_{MN}\) and \(Q^\mathrm{gauge}_M\) do not exist
and the theories are always ungauged, consistent with \eqref{eq:hgauge}.
On the other hand the possible background fluxes of higher rank field strengths
are not restricted. Similarly,
the six-dimensional \(\cN = (1,0)\) theories
cannot be deformed by \(A_0 \neq 0\) and do not feature any vector
fields in the gravity multiplet.
In principle it is possible to gauge these theories by
coupling them to vector multiplets. However, in the maximally
supersymmetric background this is forbidden due to \eqref{eq:hgauge}
and therefore also here \(Q^\mathrm{gauge}_M = 0\) holds. 
This was explicitly shown in \cite{Akyol:2010iz}.

The analysis of the two remaining supergravities in the list, the four- and five-dimensional \(\cN = 2\) theories, is slightly more involved.
Both can be deformed by \(A_0 \neq 0\) and both have one single gauge field, 
the graviphoton \(A^{\hat\alpha_2}\), in the gravity multiplet.
%For convenience let us denote is  \(\hat A_M\) with field strength \(\hat F_{MN}\). \Jnote{better notation?}
Consequently there is also only one single matrix
\(B_{\hat\alpha_2}\).
% which we denote by \(\hat B\).
As the graviphotons is an R-symmetry singlet, \(B_{\hat\alpha_2}\)
has to be proportional to the identity.
Therefore \eqref{eq:hgauge} simply reads
\begin{equation}
\cH^\mathrm{gauge}_{MN} \sim  F_{MN} A_0 \,,
\end{equation}
where $F_{MN}$ is the field strength of the graviphoton.
As a consequence, \eqref{eq:hgaugecondition} implies that 
either \(F_{MN}\)  or \(A_0\) has to vanish in the background.
%In the first case there are no background fluxes at all, while in the second case the theory is effectively ungauged since \(Q^\mathrm{gauge}_M = A_0 = 0\).
For $N=2$ theories in \(D=4\) this has been explicitly shown for
pure gauged supergravity in \cite{Caldarelli:2003pb} and for arbitrary
gauging in \cite{Hristov:2009uj}. For pure gauged supergravity in
\(D=5\) this has been obtained in \cite{Gauntlett:2003fk} and related
results for arbitrary gaugings in \cite{Bellorin:2008we}.
In contrast to their results our analysis here is completely general and does not rely on the concrete formulation of the gauged supergravities.

Let us summarize our results so far. There are two different branches of maximally supersymmetric solutions:
%\paragraph{
\begin{itemize}
\item[i)] \(A_0 \neq 0\).

In this case all background fluxes must necessarily vanish and the background space-time is \(AdS_D\) as described in section~\ref{sec:susybackgrounds}.
%\paragraph{b) \(A_0 = 0\)}

\item[ii)] \(A_0 = 0\).

In this case non-vanishing background fluxes are allowed but \(Q_M\)
vanishes in the background. As a consequence the fermionic
supersymmetry transformation \eqref{eq:gravitinovariation}
take exactly the same form as for the ungauged theory and
hence the maximally supersymmetric solutions coincide
with the solutions of the ungauged theories. 

\end{itemize}

The solutions of the ungauged theories  have been classified for 
all supergravities listed in table~\ref{tab:pformfluxes}
and this classification  can thus  be used for  case ii).
These solutions can be found in the references given in table~\ref{tab:pformfluxes}.
% For \(D=4\), \(\cN =2\) this has been done in \cite{Tod:1983pm}, for \(D=5\), \(\cN =2\) in \cite{Gauntlett:2002nw}, for \(D=6\), \(\cN = (1,0)\) in \cite{Gutowski:2003rg}, generalized to  \(\cN = (2,0)\) in \cite{Chamseddine:2003yy}, and for \(D=10, 11\) in \cite{FigueroaO'Farrill:2002ft}. \Snote{Referenzen in die Tabelle einfuegen?}
%
Let us shortly review the main results.
For vanishing \(A_0\) and \(Q_M\) the integrability condition \eqref{eq:fluxgaugeintegrability} simplifies considerably and reads
\begin{equation}\label{eq:fluxintegrability}
\frac{1}{4}R_{MNPQ} \Gamma^{PQ} \delta^i_j + 2 \left(\nabla_{[M} \cF_{N]}\right)^i_j
+ 2 \left(\cF_{[M}\right)^i_k\left(\cF_{N]}\right)^k_j = 0 \ .
\end{equation}
Expanding in powers of the \(\Gamma\)-matrices and collecting all  terms quadratic in \(\Gamma\) we observe that 
the Riemann tensor of the space-time background is expressed solely in terms of the background flux \(F^{\hat\alpha_p}\) and its derivatives.
Furthermore, all supergravities listed in table~\ref{tab:pformfluxes}
have solutions  with the property
\begin{equation}\label{eq:nablaf}
\nabla F^{\hat\alpha_p} = 0 \,.
\end{equation}
Only in the five-dimensional \(\cN = 2\) supergravity one finds
solutions of \eqref{eq:fluxintegrability}
which do not satisfy \eqref{eq:nablaf} \cite{Gauntlett:2002nw}.
In all other cases  %listed in table~\ref{tab:pformfluxes}
there are no additional solutions or in other words 
\emph{all} solutions share the property
\eqref{eq:nablaf}. 
%To prove this one has again to evaluate \eqref{eq:fluxintegrability} order by order in the \(\Gamma\)-matrices.
For these solutions
also the Riemann tensor is parallel, i.e.\ \(\nabla_M R_{NPQR} = 0\), which says that the space-time is locally symmetric.
The locally symmetric spaces with Lorentzian signature are classified 
\cite{Cahen1970, FigueroaO'Farrill:2002ft}.\footnote{They have to be 
locally isometric to 
a product of a Riemannian symmetric space times a Minkowskian, dS, 
AdS or H{\it pp}-wave geometry.}
Furthermore, in \cite{Gutowski:2003rg,Chamseddine:2003yy,FigueroaO'Farrill:2002ft}
it was shown that $F^{\hat\alpha_p}$ can be written as
\begin{equation}\label{eq:decomp}
F^{\hat\alpha_p} = v^{\hat\alpha_p} F \qquad \text{or} \qquad F^{\hat\alpha_p} = v^{\hat\alpha_p}\left(F + \ast F\right) \,,
\end{equation}
where \(v^{\hat\alpha_p}\) is constant and \(F\) is decomposable, i.e.\
 it can always be expressed as the wedge-product of \(p\) one-forms.
The second decomposition holds for a self-dual \(F^{\hat\alpha_p}\).\footnote{Notice that in \(D=4\) dimensions \(F^{\hat\alpha_2}\) itself is not necessarily decomposable.
Instead we have to split it into a complex self-dual and anti-self-dual part and use the appropriate form of the second decomposition in \eqref{eq:decomp}.}
Excluding the trivial case where \(F = 0\) and where the background is flat, there are therefore only two cases to be distinguished:
\begin{enumerate}
%\item \(F = 0\):
%This case has been already discussed in section~\ref{sec:susybackgrounds} and implies that the background space-time is flat, i.e. D-dimensional Minkowski space.
%\begin{equation}
%\cM_D = M_D \,,
%\end{equation}
% where \(M_D\) denotes D-dimensional Minkowski space-time.
\item \(F\) is \emph{not} a null form (i.e.\ $F^2\neq0$).

These are the well-known solutions of Freund-Rubin type \cite{Freund:1980xh} for which the space-time is the product of an AdS space and a sphere such that \(F\) is a top-form on one of the two factors, i.e.
\begin{equation}
\cM_D = \AdS_p \times S^{(D-p)} \qquad\text{or}\qquad \cM_D = \AdS_{(D-p)} \times S^p \,.
\end{equation}
We explicitly list all these solutions in table~\ref{tab:adsbackgrounds}. 
Notice that besides the pure \(\AdS_D\) solutions discussed in section~\ref{sec:susybackgrounds}
these are the only possible maximally supersymmetric solutions with an \(\AdS\)-factor. All other \(\AdS\) solutions in supergravity will necessarily break supersymmetry.

\item \(F\) is a null form (i.e.\ $F^2=0$).

These solutions are homogeneous \textit{pp}-waves (H\textit{pp}-waves)
first discovered by Kowalski-Glikman \cite{KowalskiGlikman:1984wv,
  KowalskiGlikman:1985im} and therefore often referred to as KG solutions.
They can be obtained from the respective \(\AdS \times S\) solutions by a Penrose limit \cite{Penrose1976, Gueven:2000ru, Blau:2002dy, Blau:2002mw}.
\end{enumerate}

As we have already mentioned above this list of solutions is exhaustive if one excludes the five-dimensional \(\cN = 2\) supergravity.
In the latter theory there can be more exotic solutions with \(F\) not parallel or decomposable and consequently also the background space-time \(\cM_D\) not locally symmetric.
These exceptional solutions are classified in \cite{Gauntlett:2002nw}
and are a G\"odel-like universe and the near-horizon limit of the
rotating BMPV  black hole \cite{Breckenridge:1996is}.%
\footnote{In \cite{Gauntlett:2002nw} three additional solutions have been found but were left unidentified, it was shown in\cite{Fiol:2003yq} that they also belong to the family of near-horizon BMPV solutions. See also \cite{Chamseddine:2003yy}.} The latter family of solutions contains the \(AdS_2 \times S^3\) and \(AdS_3 \times S^2\) solutions as special cases.
Even though there are maximally supersymmetric solutions which are not locally symmetric, they all happen to be homogeneous space-times \cite{Cahen1970, AlonsoAlberca:2002wr, Chamseddine:2003yy}.
It is also interesting to note that the maximally supersymmetric solutions of the theories with 8 real supercharges in \(D=4,5,6\) dimensions are related via dimensional reduction or oxidation \cite{LozanoTellechea:2002pn, Chamseddine:2003yy}.

%The technical reason for this can be traced down to the fact that here the term quadratic and cubic in \(\Gamma\) in \eqref{eq:fluxintegrability} can be dualized into each other by means of \eqref{}.

\begin{table}[htb]
\centering
\begin{tabular}{|c|c|c|cl|l|c|}
\hline
dim. & SUSY & q %\# real supercharges 
& \multicolumn{2}{|c|}{\(AdS \times S\)} &  H\textit{pp}-wave & others \\
\hline
\multirow{2}{*}{\(D = 11\)} & \multirow{2}{*}{\(\cN = 1\)} & \multirow{2}{*}{32} & \(\AdS_4 \times S^7\) &\multirow{2}{*}{\cite{Freund:1980xh}}& \multirow{2}{*}{\(\mathrm{KG}_{11}\) \cite{KowalskiGlikman:1984wv}} & \multirow{2}{*}{-}  \\
&&&\(\AdS_7 \times S^4\) &
%,Biran:1982eg,Pilch:1984xy} 
&& \\ [0.9ex]
\(D = 10\) & IIB & 32 & \(\AdS_5 \times S^5\)& \cite{Schwarz:1983qr,Schwarz:1983wa} & \(\mathrm{KG}_{10}\) \cite{Blau:2001ne} & - \\[0.9ex]
\multirow{2}{*}{\(D=6\)} & \(\cN = (2,0)\) & 16 & \multirow{2}{*}{\(\AdS_3 \times S^3\)} &\multirow{2}{*}{\cite{Gibbons:1994vm}}& \multirow{2}{*}{\(\mathrm{KG}_{6}\)\cite{Meessen:2001vx} } & \multirow{2}{*}{-} \\
& \(\cN = (1,0)\) & 8 & && & \\[0.9ex]
\multirow{2}{*}{\(D = 5\)} & \multirow{2}{*}{\(\cN = 2\)} & \multirow{2}{*}{8} & \(\AdS_2 \times S^3\)&\multirow{2}{*}{\cite{Gibbons:1994vm,Chamseddine:1996pi}} & \multirow{2}{*}{\(\mathrm{KG}_{5}\) \cite{Meessen:2001vx}} & G\"odel-like \cite{Gauntlett:2002nw}, \\
&&& \(\AdS_3 \times S^2\)  &&& NH-BMPV \cite{Cvetic:1998xh,Gauntlett:1998fz} \\[0.9ex]
\(D = 4\) & \(\cN = 2\) & 8 & \(\AdS_2 \times S^2\) &\cite{Bertotti:1959pf,Robinson:1959ev}& \(\mathrm{KG}_{4}\) \cite{KowalskiGlikman:1985im} & - \\
\hline
\end{tabular}
\caption{All possible maximally supersymmetric solutions with
  non-trivial flux; $q$ denotes the number of real supercharges, cf.\
\cite{AlonsoAlberca:2002dw}.}
\label{tab:adsbackgrounds}
\end{table}

%%%%%%%%%%%%%%%%%%%%%%%%%%%%%%%%%%%%%%%%%%%%%%%%%%%%%%%%%%%
%\newpage

\section{Conclusions}

%Let us briefly conclude.
In this paper we studied maximally supersymmetric solutions of all  supergravities in  space-time dimensions $3\le D\le 11$ -- including gauged supergravities
as well as supergravities with background fluxes.
%Hereby we tried to work in a generic framework that is hiding away as much as possible the dimension and supersymmetry dependent details.
%
We found that the maximally supersymmetric solutions generically split into three separate classes.
First of all there are the ungauged and undeformed 
supergravities without fluxes and a
$D$-dimensional Minkowskian background.  
The second class of solutions consists of backgrounds without fluxes but 
the supergravity is gauged or otherwise deformed.
In this case the Killing spinor equations are %exceedingly 
straightforward to integrate, implying that the space-time is maximally symmetric and therefore either again Minkowskian or \(\AdS_D\).
There are however certain algebraic conditions \eqref{AdSsolution} 
which the fermionic shift-matrices \(A_0\), \(A_1\) and \(A_2\) have to 
satisfy and which restrict the possible gaugings or deformations.

The third class of solutions has non-trivial
background  fluxes. This requires that
all shift matrices \(A_0, A_1, A_2\) vanish and the R-symmetry
connection has no background value. It implies 
that the fermionic supersymmetry variations take exactly the same form 
as for the corresponding ungauged theories.
Moreover, this class of solutions can only exist if either the gravitational multiplet has no spin-$\frac12$ fermions or the theory is chiral. 
This selects among all supergravities the ones listed in
table~\ref{tab:pformfluxes} and in addition selects the possible
fluxes.
Using the correspondence with the ungauged theories we argued 
that for all these theories all solutions are known and classified;
we list them  in table~\ref{tab:adsbackgrounds}. One aspect of our
analysis was to  show that this list is exhaustive.

Of course in certain cases solutions from different classes can be related to each other.
The solutions of the form \(\AdS_{d} \times S^{(D-d)}\) -- which might
arise in \emph{ungauged} theories -- can be truncated to an effective
\(d\)-dimensional description in terms of a \emph{gauged} supergravity
with an \(\AdS_d\) background.
The gauge group in this case is  \(SO(D-d+1)\), i.e.\ 
the isometry group of the sphere \(S^{(D-d)}\).
At the same time not every AdS-solution of a gauged supergravity can be obtained from a sphere compactification. 
 
As a technical by-product of our analysis we derived the general formula \eqref{eq:maxsusyrgenerators} for the gauging of the  R-symmetry in maximally supersymmetric solutions in dimensions \(D \geq 4\).
It shows that in the background the R-symmetry can only be gauged by vector fields from the gravitational multiplet (i.e.\ graviphotons) and that this gauging is completely fixed by the first shift matrix \(A_0\) and therefore uniquely determined.

%In this paper we studied AdS backgrounds of all supergravities
%in  $3\le D\le 11$ space-time dimensions.
%We exclusivly focused on  AdS backgrounds which preserve all supercharges
%of a given supergravity. We found that apart from flat space solutions
%only backgrounds of the form
%\begin{equation}
%\cM_D = \AdS_D \quad \textrm{or}\quad 
%\cM_D = \AdS_d \times S^{(D-d)} \ ,\qquad 1\le d \leq D-1\ 
%\end{equation} 
%are possible. 
%The second class of backgrounds 
%only exist if either
%the gravitational multiplet has no spin-$\frac12$ fermions or the theory is chiral. 
%This selects among all supergravities the ones given in Table~\ref{tab:pformfluxes} with specific fluxes. We found that
%a top-form flux on at least one of the 
%two factors has to be non-trivial which, combined with Table~\ref{tab:pformfluxes}, selects the small list of backgrounds given in Table~\ref{tab:adsbackgrounds}.
%All of these backgrounds were previously known but our analysis shows that this list is exhaustive.

%%%%%%%%%%%%%%%%%%%%%%%%%%%%%%%%%%%%
\section*{Acknowledgments}
%%%%%%%%%%%%%%%%%%%%%%%%%%%%%%%%%%%%

This work was supported by the German Science Foundation (DFG) under
the Collaborative Research Center (SFB) 676 ``Particles, Strings and the Early
Universe'' and the Research Training Group (RTG) 1670 ``Mathematics
inspired by String Theory and Quantum Field Theory''.

We have benefited from conversations and correspondence with  
Eric Bergshoeff, Constantin Muranaka, Hagen Triendl and  Dan Waldram.

%%%%%%%%%%%%%%%%%%%%%%%%%%%%%%%%%%%%%%%%%%%%%%%%%%%%%%%%%%%
\newpage

\appendix
\noindent
{\bf\Huge Appendix}

\section{$\Gamma$-matrix conventions} \label{app:gamma}

In this appendix we collect some useful $\Gamma$-matrix identities 
used throughout the paper.
%and carry out some computations in detail which we omitted in the main part.
We mainly follow the definitions and conventions of \cite{Freedman:2012zz}.
The \(\Gamma^M\) are defined via their anti-commutation relation
%\Snote{Flat metric \(\eta^{MN}\) vs. curved \(g^{MN}\)?}
\begin{equation}\label{eq:gammadef}
\Gamma^M \Gamma^N + \Gamma^N \Gamma^M = 2 g^{MN} \id \,.
\end{equation}
Frequently in the main text their antisymmetric products 
appear and we abbreviate
\begin{equation}\label{eq:gammaprod}
\Gamma^{M_1\dots M_p} = \Gamma^{[M_1} \dots \Gamma^{M_p]} \,,
\end{equation}
where the antisymmetrization \([\dots]\) is with total weight 1, i.e. \(\Gamma^{MN} = \frac{1}{2}\left(\Gamma^M\Gamma^N - \Gamma^N \Gamma^M\right)\).
In even dimensions \(D = 2m\) we additionally have the chirality operator 
\(\Gamma_\ast\) defined by 
\begin{equation}\label{eq:gamma5}
\Gamma_\ast = (-i)^{m+1} \Gamma_0 \Gamma_1 \dots \Gamma_{D-1} \,.
\end{equation}
From its definition one infers \cite{Freedman:2012zz}
%and the gamma matrices satisfy \Snote{Directly copied from Freedman, van Proeyen...}
\begin{equation}\label{eq:gammahodgeeven}
\Gamma^{M_1 \dots M_p}\, \Gamma_\ast = -(-i)^{m+1} \frac{1}{(D-p)!}\, {\epsilon^{M_p \dots M_1}}_{N_1 \dots N_{D-p}}\, \Gamma^{N_1 \dots N_{D-p}} \,,
\end{equation}
while in odd dimensions \(D = 2m +1\) one has instead
\begin{equation}\label{eq:gammahodgeodd}
\Gamma^{M_1 \dots M_p} = i^{m+1} \frac{1}{(D-p)!} {\epsilon^{M_1 \dots M_p}}_{N_{D-p} \dots N_1} \Gamma^{N_1 \dots N_{D-p}} \,.
\end{equation}
In even dimensions all anti-symmetric products $\Gamma^{M_1\dots M_p}$
are linearly independent whereas in odd dimensions this only holds for \(p \leq m\) due to \eqref{eq:gammahodgeodd}. 
Moreover we denote the contraction with $\Gamma$-matrices by a dot ``\(\cdot\)", i.e. for a \(p\)-form~\(F\) we define
\begin{equation}
F \cdot \Gamma = F_{M_1 \dots M_p}\, \Gamma^{M_1 \dots M_p} \,.
\end{equation}

\section{Analysis of the integrability condition}\label{app:integrability}

In this appendix we analyze the integrability condition \eqref{eq:fluxgaugeintegrability} and argue that for all the theories listed in table~\ref{tab:pformfluxes} the term \(\cH_{MN}\) can be the only term at zeroth order in the \(\Gamma\)-matrices and has therefore to vanish in a maximally supersymmetric background.

Let us first note that all the theories in table~\ref{tab:pformfluxes} only allow for background fluxes \(F^{\hat\alpha_p}\) for one particular value of \(p\), so the expression \eqref{cFdef} for \(\cF_M\) simplifies as we do not have to sum over different values for \(p\).
We want to inspect \eqref{eq:fluxgaugeintegrability} term by term.
While the Riemann tensor \(R_{MNPQ}\) enters only at the quadratic order in \(\Gamma\), also the third term \(\left(\D_{[M} \cF_{N]} + \D_{[M} A_0 \Gamma_{N]}\right)\) cannot contain any terms at zeroth order in \(\Gamma\) as can be directly seen from \eqref{cFdef} and \eqref{eq:T} with \(p > 1\).
    To analyze the remaining term in \eqref{eq:fluxgaugeintegrability} we notice that this term can only produce something of vanishing order in \(\Gamma\) from the anti-commutator of two equal powers of \(\Gamma\)-matrices, i.e.
\begin{equation}\label{eq:multgammaanticom}
\left\{\Gamma^{M_1\dots M_r},\Gamma_{N_1\dots N_r}\right\} = p!\, \delta^{[M_1}_{N_r} \dots \delta^{M_r]}_{N_1} + \dots \,,
\end{equation}
where the dots denote terms of higher order in \(\Gamma\).
On the hand the corresponding commutator  yields at least a term
quadratic
in \(\Gamma\) and also the (anti-)commutator of two different powers of \(\Gamma\)-matrices cannot give anything at zeroth order.
With this knowledge we can finally compute the last term in \eqref{eq:fluxgaugeintegrability} to find
%finally see that only the commutator \(\left[\cF_M,\cF_N\right]\) could give something at zeroth order.
%The relation \eqref{eq:multgammaanticom} yields with \eqref{cFdef}
\begin{equation}\begin{aligned}\label{eq:lastterm}
\Bigl(\left(\cF_M + A_0 \Gamma_M\right)&\left(\cF_N + A_0 \Gamma_N\right) - (M \leftrightarrow N)\Bigr) = \\
&= \bigl[\cF_M,\cF_N\bigr] + \bigl[\cF_M, A_0 \Gamma_N\bigr] - A_0 \bigl[\cF_N, A_0 \Gamma_M\bigr] + 2 A_0 A_0 \Gamma_{MN} \\
&= \frac{1}{2}(p-1)!\left(\beta_{(p)}^2-p^2\right)\bigl[B_{\hat\alpha_p},B_{\hat\beta_p}\bigr] F^{\hat\alpha_p}_{MP_1\dots P_{p-1}} F^{\hat\beta_p\,P_{p-1}\dots P_p}_N \\
&\qquad+ \delta_{p,2}\, 4(D-3) \bigl[B_{\hat\alpha_2}, A_0\bigr] F^{\hat\alpha_2}_{MN} + \dots \,,
\end{aligned}\end{equation}
where we suppressed the indices \((i,j,\dots)\) and the dots denote again higher order terms.
For the computation of the commutator \(\bigl[\cF_M,\cF_N\bigr]\) we
used \eqref{cFdef}, \eqref{eq:multgammaanticom} and 
\begin{equation}\begin{aligned}
&\bigl[B_{\hat\alpha_p} \Gamma^{M_1\dots M_r}, B_{\hat\beta_p}\Gamma^{N_1\dots N_r} \bigr] \\
&\qquad= \frac{1}{2} \left(\bigl[B_{\hat\alpha_p},B_{\hat\beta_p}\bigr] \left\{\Gamma^{M_1\dots M_r},\Gamma^{N_1\dots N_r}\right\} + \bigl\{B_{\hat\alpha_p},B_{\hat\beta_p}\bigr\} \left[\Gamma^{M_1\dots M_r},\Gamma^{N_1\dots N_r}\right] \right) \,.
\end{aligned}\end{equation}
For all the theories where \(\hat\alpha_p\) can take only one possible value the commutator \(\bigl[B_{\hat\alpha_p},B_{\hat\beta_p}\bigr]\) on the right hand side of \eqref{eq:lastterm} clearly vanishes.
Moreover in this case \(B_{\hat\alpha_p}\) is proportional to the unit matrix, therefore also the second commutator \(\bigl[B_{\hat\alpha_2}, A_0\bigr]\) vanishes.
The only theory in table~\ref{tab:pformfluxes} for which \(\hat\alpha_p\) can take multiple values is the six-dimensional \(\cN = (2,0)\) theory.
But here \(p = D/2 = 3\) and hence using \eqref{eq:beta} we have 
\(\beta_{(p)} = p\) so that also in this case 
the terms at zeroth order in \(\Gamma\) vanish.

It remains to check that in odd dimensions \(D\) there are also no terms of order \(D\) in \(\Gamma\).
These could be dualized into zero order terms using \eqref{eq:gammahodgeodd}.
Since we can restrict the analysis to \(p < \frac{D}{2}\) it is clear
that such terms cannot arise from \(\D_{[M} \cF_{N]}\) or
\(\bigl[\cF_M A_0, \Gamma_N\bigr]\)
as can be seen from the definition \eqref{cFdef}.
The commutator \(\bigl[\cF_M,\cF_N\bigr]\) can however produce only terms of even order in \(\Gamma\).

\section{The gauged R-symmetry connection}\label{app:rconnection}

In this appendix we show that in a maximally supersymmetric background %of a gauged supergravity in \(D > 3\) 
the R-symmetry can only be gauged by vector fields from the gravity multiplet (graviphotons).% or not at all.
\footnote{See also the discussion below equation \eqref{eq:rgenerators}.}
To be more specific, we will show that a non-vanishing background value for those moment maps \(t_{\tilde\alpha_2}\) from \eqref{eq:rgenerators}
which couple to the 
%Gauging the R-symmetry by 
vector fields from other
multiplets 
(i.e.\ vector multiplets) is not compatible with unbroken supersymmetry.
We should stress that this is only a statement about the maximally supersymmetric background, at an arbitrary point in field space these restrictions on the gaugings do not necessarily need to be satisfied.
In other words, a theory in which there are gauge fields that do not  belong to the gravity multiplet might still admit maximally supersymmetric vacua.
Note that the following analysis does not rely 
on any specific formulation of a gauged supergravity and is valid in any dimension \(D > 3\).

%In this appendix we show that a gauged supergravity in \(D > 3\) does admit
%maximally supersymmetric solutions  if the R-symmetry is gauged
% only by
%vector fields from the gravity multiplet (graviphotons) or not gauged at all.
%In other words gauging the R-symmetry by vector fields from other
%multiplets 
%(for example  vector multiplets) necessarily breaks supersymmetry.
%Note that the following analysis does  not rely 
%on any specific formulation of a gauged supergravity.

As explained in section~\ref{sec:prelim} the R-symmetry connection \(Q_M\) generically splits into a pure scalar dependent part \(Q_M^\mathrm{scalar}\) and a gauge field depend part \(Q_M^\mathrm{gauge}\).
The latter can be expressed in terms of the gauge fields \(A^{\alpha_2}\) as
\begin{equation}
\left(Q^\mathrm{gauge}_M\right)^i_j = A_M^{\alpha_2} \left(t_{\alpha_2}\right)^i_j \,.
\end{equation}
%where the \(t_{\alpha_2}\) generate the gauged subgroup of the R-symmetry group.
Equivalently the corresponding part of the field strength \(\cH_{MN}\) reads
$\cH_{MN}^\mathrm{gauge} = F_{MN}^{\alpha_2} t_{\alpha_2}$.
In the following we determine the explicit form of the matrices \(t_{\alpha_2}\) 
in the maximally supersymmetric background,
%such that the gauging allows for solutions with unbroken supersymmetry,
i.e. for vanishing fermionic shift matrices \(A_1\) and \(A_2\).
 
The field strength \(\cH_{MN}\) enters the supersymmetry variation of the Lagrangian via the kinetic term for the gravitini which always takes the form
\begin{equation}
e^{-1} \cL_{\bar\psi\partial\psi} = - \tfrac{1}{2} \bar \psi_{iM} \Gamma^{MNP} \D_N \psi^i_P \,. 
\end{equation}
Inserting into this the supersymmetry variation \eqref{eq:gravitinovariation} of \(\psi^i_M\) produces a term of the form
\begin{equation}\label{eq:psikinvariation}
e^{-1} \delta \cL_{\bar\psi\partial\psi} =
\tfrac{1}{2}\left(\cH^\mathrm{gauge}_{MN}\right)^i_j \bar \psi_{iP}
\Gamma^{MNP} \epsilon^j +\ldots \,.
\end{equation}
To read off \(\cH_{MN}\) we collect all possible terms which produce similar
terms under a supersymmetry transformation.
If we demand \(A_1 = A_2 = 0\) these are given by
\begin{equation}\begin{aligned}\label{Lint}
e^{-1}\cL_{\bar\psi\psi} &= \tfrac{1}{2} d_0 A^i_{0\,j} \bar\psi_{iM} \Gamma^{MN} \psi^j_N \,,\\
e^{-1}\cL_{F\bar\psi\psi} &=  \tfrac{1}{2} e_0 F^{\hat\alpha_2}_{MN} \left(B_{\hat\alpha_2}\right)^i_j \bar \psi_i^P \Gamma_{[P}\Gamma^{MN}\Gamma_{R]} \psi^{jR} \,,
\end{aligned}\end{equation}
where the numerical constants \(d_0\) and \(e_0\) are fixed by supersymmetry and can be determined to take the values
\begin{equation}
d_0 = -e_0 = (D-2) \,.
\end{equation}
From \eqref{eq:gravitinovariation} it follows that the supersymmetry variations of \eqref{Lint} contain precisely terms of the required form
\begin{equation}
e^{-1}\delta \cL_{\bar\psi\psi} = d_0 F^{\hat\alpha_2}_{MN} A^i_{0\,j}
\left(B_{\hat\alpha_2}\right)^j_k \bar\psi^P_i \left(-(D-3)
   {\Gamma^{MN}}_P + 2 \delta_P^{[M}\Gamma^{N]}\right) \epsilon^k
+\ldots \,,
\end{equation}
and
\begin{equation}
e^{-1} \delta \cL_{F\bar\psi\psi} = e_0 F^{\hat\alpha_2}_{MN}  \left(B_{\hat\alpha_2}\right)^i_j A^j_{0\,k} \bar\psi^P_i \left((D-3) {\Gamma^{MN}}_P + 2 \delta_P^{[M}\Gamma^{N]}\right) \epsilon^k +\ldots\,.
\end{equation}
The terms cubic in \(\Gamma\)-matrices have to cancel \eqref{eq:psikinvariation}, so we finally determine 
\begin{equation}\label{eq:maxsusyrgenerators}
t_{\hat\alpha_2} = 2 (D-2) (D-3) \bigl\{A_0, B_{\hat\alpha_2}\bigr\} \,,\qquad t_{\tilde\alpha_2} = 0 \,.
\end{equation}
This  means in particular that in a maximally supersymmetric background the R-symmetry can only be gauged by the graviphotons \(A^{\hat\alpha_2}\), but not by gauge fields \(A^{\tilde\alpha_2}\) in additional vector multiplets.

Finally we want to argue that the matrices \(t_{\hat\alpha_2}\) satisfying \eqref{eq:maxsusyrgenerators} span a Lie-subalgebra of the R-symmetry algebra \(\mathfrak{g}_R\), i.e. that they close with respect to the Lie-bracket.
Using the fact that we always need \(A_0^2 \sim \mathbb{1}\) in a maximally supersymmetric background (according to \eqref{AdSsolution}), we immediately find that
\begin{equation}\label{eq:A0invariance}
\left[t_{\hat\alpha_2}, A_0\right] = 0 \,,
\end{equation}
i.e. that \(A_0\) is invariant under the adjoint action of the \(t_{\hat\alpha_2}\).
To proceed, let us denote the generators of \(\g_R\) by \(T_A\), \(A = 1, \dots, \dim(\g_R)\).
Now invariance of the gravitino variations \eqref{eq:gravitinovariation} under R-symmetry transformations requires all the \(B\)-matrices to be \(\g_R\) invariant, in the sense that
\begin{equation}\label{eq:Binvariance}
{\left(T_A\right)_{\hat\alpha_2}}^{\hat\beta_2} B_{\hat\beta_2} - \left[T_A, B_{\hat\alpha_2}\right] = 0 \,.
\end{equation}
Here \({\left(T_A\right)_{\hat\alpha_2}}^{\hat\beta_2}\) and \({\left(T_A\right)_i}^j\) denote the R-symmetry generators in the representations of the graviphotons and the gravitini respectively.
So loosely speaking the \(B\)-matrices ``translate'' between different representations of \(\g_R\).
As \(t_{\alpha_2} \in \g_R\) we can always find (generically scalar dependent) matrices \({\Theta_{\alpha_2}}^A\) such that
\begin{equation}
t_{\alpha_2} = {\Theta_{\alpha_2}}^A T_A \,.
\end{equation}
Using this information it follows from \eqref{eq:A0invariance} and \eqref{eq:Binvariance} that
\begin{equation}\begin{aligned}
\bigl[t_{\hat\alpha_2}, t_{\hat\beta_2}\bigr] &= 2(D-2)(D-3) \left\{A_0, \bigl[t_{\hat\alpha_2}, B_{\hat\beta_2}\bigr]\right\} \\
&= 2(D-2)(D-3) \left\{A_0, {\left(t_{\hat\alpha_2}\right)_{\hat\beta_2}}^{\hat\gamma_2} B_{\hat\gamma_2}\right\} \\
&= {\left(t_{\hat\alpha_2}\right)_{\hat\beta_2}}^{\hat\gamma_2} \, t_{\hat\gamma_2} \,,
\end{aligned}\end{equation}
where we have introduced \({\left(t_{\hat\alpha_2}\right)_{\hat\beta_2}}^{\hat\gamma_2} = {\Theta_{\hat\alpha_2}}^A {\left(T_A\right)_{\hat\beta_2}}^{\hat\gamma_2}\).

%%%%%%%%%%%%%%%%%%%%%%%%%%%%%%%%%%%%%%%%%%%%%%%%%%%%%%%%%%%
%\newpage

\providecommand{\href}[2]{#2}\begingroup\raggedright\endgroup


\begin{thebibliography}{10}

%\cite{Tod:1983pm}
\bibitem{Tod:1983pm}
  K.~P.~Tod,
  ``All Metrics Admitting Supercovariantly Constant Spinors,''
  Phys.\ Lett.\ B {\bf 121} (1983) 241.
  %%CITATION = doi:10.1016/0370-2693(83)90797-9;%%
  %239 citations counted in INSPIRE as of 11 Jul 2016

%\cite{Behrndt:1997ny}
\bibitem{Behrndt:1997ny}
  K.~Behrndt, D.~Lust and W.~A.~Sabra,
  ``Stationary solutions of N=2 supergravity,''
  Nucl.\ Phys.\ B {\bf 510} (1998) 264
  %doi:10.1016/S0550-3213(97)00633-0, 10.1016/S0550-3213(98)81014-6
  [hep-th/9705169].
  %%CITATION = doi:10.1016/S0550-3213(97)00633-0, 10.1016/S0550-3213(98)81014-6;%%
  %189 citations counted in INSPIRE as of 17 Nov 2016

%\cite{Gauntlett:2002sc}
\bibitem{Gauntlett:2002sc}
  J.~P.~Gauntlett, D.~Martelli, S.~Pakis and D.~Waldram,
  ``G structures and wrapped NS5-branes,''
  Commun.\ Math.\ Phys.\  {\bf 247} (2004) 421
  %doi:10.1007/s00220-004-1066-y
  [hep-th/0205050].
  %%CITATION = doi:10.1007/s00220-004-1066-y;%%
  %256 citations counted in INSPIRE as of 17 Nov 2016

%\cite{Gauntlett:2002nw}
\bibitem{Gauntlett:2002nw}
  J.~P.~Gauntlett, J.~B.~Gutowski, C.~M.~Hull, S.~Pakis and H.~S.~Reall,
  ``All supersymmetric solutions of minimal supergravity in five- dimensions,''
  Class.\ Quant.\ Grav.\  {\bf 20} (2003) 4587
   [hep-th/0209114].
  %%CITATION = doi:10.1088/0264-9381/20/21/005;%%
  %462 citations counted in INSPIRE as of 23 Jun 2016

%\cite{FigueroaO'Farrill:2002ft}
\bibitem{FigueroaO'Farrill:2002ft}
  J.~M.~Figueroa-O'Farrill and G.~Papadopoulos,
  ``Maximally supersymmetric solutions of ten-dimensional and eleven-dimensional supergravities,''
  JHEP {\bf 0303} (2003) 048
%  doi:10.1088/1126-6708/2003/03/048
  [hep-th/0211089].
  %%CITATION = doi:10.1088/1126-6708/2003/03/048;%%
  %133 citations counted in INSPIRE as of 20 Feb 2016

%\cite{Gauntlett:2002fz}
\bibitem{Gauntlett:2002fz}
  J.~P.~Gauntlett and S.~Pakis,
  ``The Geometry of D = 11 killing spinors,''
  JHEP {\bf 0304} (2003) 039
  %doi:10.1088/1126-6708/2003/04/039
  [hep-th/0212008].
  %%CITATION = doi:10.1088/1126-6708/2003/04/039;%%
  %260 citations counted in INSPIRE as of 17 Nov 2016

%\cite{Gauntlett:2003cy}
\bibitem{Gauntlett:2003cy}
  J.~P.~Gauntlett, D.~Martelli and D.~Waldram,
  ``Superstrings with intrinsic torsion,''
  Phys.\ Rev.\ D {\bf 69} (2004) 086002
  %doi:10.1103/PhysRevD.69.086002
  [hep-th/0302158].
  %%CITATION = doi:10.1103/PhysRevD.69.086002;%%
  %273 citations counted in INSPIRE as of 17 Nov 2016

%\cite{Gutowski:2003rg}
\bibitem{Gutowski:2003rg}
  J.~B.~Gutowski, D.~Martelli and H.~S.~Reall,
  ``All Supersymmetric solutions of minimal supergravity in six- dimensions,''
  Class.\ Quant.\ Grav.\  {\bf 20} (2003) 5049
  [hep-th/0306235].
  %%CITATION = doi:10.1088/0264-9381/20/23/008;%%
  %153 citations counted in INSPIRE as of 23 Jun 2016

%\cite{Chamseddine:2003yy}
\bibitem{Chamseddine:2003yy}
  A.~Chamseddine, J.~M.~Figueroa-O'Farrill and W.~Sabra,
  ``Supergravity vacua and Lorentzian Lie groups,''
  hep-th/0306278.
  %%CITATION = HEP-TH/0306278;%%
  %27 citations counted in INSPIRE as of 22 Jun 2016

%\cite{Gauntlett:2003wb}
\bibitem{Gauntlett:2003wb}
  J.~P.~Gauntlett, J.~B.~Gutowski and S.~Pakis,
  ``The Geometry of D = 11 null Killing spinors,''
  JHEP {\bf 0312} (2003) 049
  %doi:10.1088/1126-6708/2003/12/049
  [hep-th/0311112].
  %%CITATION = doi:10.1088/1126-6708/2003/12/049;%%
  %114 citations counted in INSPIRE as of 17 Nov 2016

%\cite{Gauntlett:2004zh}
\bibitem{Gauntlett:2004zh}
  J.~P.~Gauntlett, D.~Martelli, J.~Sparks and D.~Waldram,
  ``Supersymmetric AdS(5) solutions of M theory,''
  Class.\ Quant.\ Grav.\  {\bf 21} (2004) 4335
  %doi:10.1088/0264-9381/21/18/005
  [hep-th/0402153].
  %%CITATION = doi:10.1088/0264-9381/21/18/005;%%
  %245 citations counted in INSPIRE as of 17 Nov 2016

%\cite{Dall'Agata:2004dk}
\bibitem{Dall'Agata:2004dk}
  G.~Dall'Agata,
  ``On supersymmetric solutions of type IIB supergravity with general fluxes,''
  Nucl.\ Phys.\ B {\bf 695} (2004) 243
  %doi:10.1016/j.nuclphysb.2004.06.037
  [hep-th/0403220].
  %%CITATION = doi:10.1016/j.nuclphysb.2004.06.037;%%
  %76 citations counted in INSPIRE as of 17 Nov 2016

%\cite{Cariglia:2004qi}
\bibitem{Cariglia:2004qi}
  M.~Cariglia and O.~A.~P.~Mac Conamhna,
  ``Timelike Killing spinors in seven dimensions,''
  Phys.\ Rev.\ D {\bf 70} (2004) 125009
  %doi:10.1103/PhysRevD.70.125009
  [hep-th/0407127].
  %%CITATION = doi:10.1103/PhysRevD.70.125009;%%
  %27 citations counted in INSPIRE as of 17 Nov 2016

%\cite{Gauntlett:2004qy}
\bibitem{Gauntlett:2004qy}
  J.~P.~Gauntlett and J.~B.~Gutowski,
  ``General concentric black rings,''
  Phys.\ Rev.\ D {\bf 71} (2005) 045002
  %doi:10.1103/PhysRevD.71.045002
  [hep-th/0408122].
  %%CITATION = doi:10.1103/PhysRevD.71.045002;%%
  %204 citations counted in INSPIRE as of 16 Dec 2016

%\cite{Gillard:2004xq}
\bibitem{Gillard:2004xq}
  J.~Gillard, U.~Gran and G.~Papadopoulos,
  ``The Spinorial geometry of supersymmetric backgrounds,''
  Class.\ Quant.\ Grav.\  {\bf 22} (2005) 1033
  %doi:10.1088/0264-9381/22/6/009
  [hep-th/0410155].
  %%CITATION = doi:10.1088/0264-9381/22/6/009;%%
  %107 citations counted in INSPIRE as of 17 Nov 2016

%\cite{Gauntlett:2004hs}
\bibitem{Gauntlett:2004hs}
  J.~P.~Gauntlett, D.~Martelli, J.~Sparks and D.~Waldram,
  ``Supersymmetric AdS backgrounds in string and M-theory,''
  IRMA Lect.\ Math.\ Theor.\ Phys.\  {\bf 8} (2005) 217
  [hep-th/0411194].
  %%CITATION = HEP-TH/0411194;%%
  %47 citations counted in INSPIRE as of 17 Nov 2016

%\cite{Gran:2005wn}
\bibitem{Gran:2005wn}
  U.~Gran, J.~Gutowski and G.~Papadopoulos,
  ``The Spinorial geometry of supersymmetric IIb backgrounds,''
  Class.\ Quant.\ Grav.\  {\bf 22} (2005) 2453
  %doi:10.1088/0264-9381/22/12/010
  [hep-th/0501177].
  %%CITATION = doi:10.1088/0264-9381/22/12/010;%%
  %72 citations counted in INSPIRE as of 17 Nov 2016

%\cite{Gutowski:2005id}
\bibitem{Gutowski:2005id}
  J.~B.~Gutowski and W.~Sabra,
  ``General supersymmetric solutions of five-dimensional supergravity,''
  JHEP {\bf 0510} (2005) 039
  %doi:10.1088/1126-6708/2005/10/039
  [hep-th/0505185].
  %%CITATION = doi:10.1088/1126-6708/2005/10/039;%%
  %54 citations counted in INSPIRE as of 17 Nov 2016

%\cite{Bellorin:2005zc}
\bibitem{Bellorin:2005zc}
  J.~Bellorin and T.~Ortin,
  ``All the supersymmetric configurations of N=4, d=4 supergravity,''
  Nucl.\ Phys.\ B {\bf 726} (2005) 171
  %doi:10.1016/j.nuclphysb.2005.07.020
  [hep-th/0506056].
  %%CITATION = doi:10.1016/j.nuclphysb.2005.07.020;%%
  %32 citations counted in INSPIRE as of 17 Nov 2016

%\cite{Ishino:2005ru}
\bibitem{Ishino:2005ru}
  T.~Ishino, H.~Kodama and N.~Ohta,
  ``Time-dependent solutions with null killing spinor in M-theory and superstrings,''
  Phys.\ Lett.\ B {\bf 631} (2005) 68
  %doi:10.1016/j.physletb.2005.09.080
  [hep-th/0509173].
  %%CITATION = doi:10.1016/j.physletb.2005.09.080;%%
  %56 citations counted in INSPIRE as of 17 Nov 2016

%\cite{Meessen:2006tu}
\bibitem{Meessen:2006tu}
  P.~Meessen and T.~Ortin,
  ``The Supersymmetric configurations of N=2, D=4 supergravity coupled to vector supermultiplets,''
  Nucl.\ Phys.\ B {\bf 749} (2006) 291
  %doi:10.1016/j.nuclphysb.2006.05.025
  [hep-th/0603099].
  %%CITATION = doi:10.1016/j.nuclphysb.2006.05.025;%%
  %75 citations counted in INSPIRE as of 17 Nov 2016

%\cite{Huebscher:2006mr}
\bibitem{Huebscher:2006mr}
  M.~Huebscher, P.~Meessen and T.~Ortin,
  ``Supersymmetric solutions of N=2 D=4 sugra: The Whole ungauged shebang,''
  Nucl.\ Phys.\ B {\bf 759} (2006) 228
  %doi:10.1016/j.nuclphysb.2006.10.004
  [hep-th/0606281].
  %%CITATION = doi:10.1016/j.nuclphysb.2006.10.004;%%
  %48 citations counted in INSPIRE as of 17 Nov 2016

%\cite{Bellorin:2006yr}
\bibitem{Bellorin:2006yr}
  J.~Bellorin, P.~Meessen and T.~Ortin,
  ``All the supersymmetric solutions of N=1,d=5 ungauged supergravity,''
  JHEP {\bf 0701} (2007) 020
  %doi:10.1088/1126-6708/2007/01/020
  [hep-th/0610196].
  %%CITATION = doi:10.1088/1126-6708/2007/01/020;%%
  %46 citations counted in INSPIRE as of 17 Nov 2016

%\cite{Cacciatori:2007vn}
\bibitem{Cacciatori:2007vn}
  S.~L.~Cacciatori, M.~M.~Caldarelli, D.~Klemm, D.~S.~Mansi and D.~Roest,
  ``Geometry of four-dimensional Killing spinors,''
  JHEP {\bf 0707} (2007) 046
  %doi:10.1088/1126-6708/2007/07/046
  [arXiv:0704.0247 [hep-th]].
  %%CITATION = doi:10.1088/1126-6708/2007/07/046;%%
  %27 citations counted in INSPIRE as of 17 Nov 2016

%\cite{Meessen:2010fh}
\bibitem{Meessen:2010fh}
  P.~Meessen, T.~Ortin and S.~Vaula,
  ``All the timelike supersymmetric solutions of all ungauged d=4 supergravities,''
  JHEP {\bf 1011} (2010) 072
  %doi:10.1007/JHEP11(2010)072
  [arXiv:1006.0239 [hep-th]].
  %%CITATION = doi:10.1007/JHEP11(2010)072;%%
  %18 citations counted in INSPIRE as of 17 Nov 2016
  
 %\cite{Akyol:2010iz}
\bibitem{Akyol:2010iz}
  M.~Akyol and G.~Papadopoulos,
  ``Spinorial geometry and Killing spinor equations of 6-D supergravity,''
  Class.\ Quant.\ Grav.\  {\bf 28} (2011) 105001
  [arXiv:1010.2632 [hep-th]].
  %%CITATION = doi:10.1088/0264-9381/28/10/105001;%%
  %5 citations counted in INSPIRE as of 12 Jul 2016 

%%\cite{Gauntlett:2005bn}
%\bibitem{Gauntlett:2005bn}
%  J.~P.~Gauntlett,
%  ``Classifying supergravity solutions,''
%  Fortsch.\ Phys.\  {\bf 53} (2005) 468
%  [hep-th/0501229].
%  %%CITATION = doi:10.1002/prop.200510206;%%
%  %19 citations counted in INSPIRE as of 25 Jul 2016

%\cite{Gauntlett:2003fk}
\bibitem{Gauntlett:2003fk}
  J.~P.~Gauntlett and J.~B.~Gutowski,
  ``All supersymmetric solutions of minimal gauged supergravity in five-dimensions,''
  Phys.\ Rev.\ D {\bf 68} (2003) 105009
   Erratum: [Phys.\ Rev.\ D {\bf 70} (2004) 089901]
  [hep-th/0304064].
  %%CITATION = doi:10.1103/PhysRevD.70.089901, 10.1103/PhysRevD.68.105009;%%
  %152 citations counted in INSPIRE as of 23 Jun 2016

%\cite{Caldarelli:2003pb}
\bibitem{Caldarelli:2003pb}
  M.~M.~Caldarelli and D.~Klemm,
  ``All supersymmetric solutions of N=2, D = 4 gauged supergravity,''
  JHEP {\bf 0309} (2003) 019
  [hep-th/0307022].
  %%CITATION = doi:10.1088/1126-6708/2003/09/019;%%
  %89 citations counted in INSPIRE as of 23 Jun 2016

%\cite{Gutowski:2004yv}
\bibitem{Gutowski:2004yv}
  J.~B.~Gutowski and H.~S.~Reall,
  ``General supersymmetric AdS(5) black holes,''
  JHEP {\bf 0404} (2004) 048
  %doi:10.1088/1126-6708/2004/04/048
  [hep-th/0401129].
  %%CITATION = doi:10.1088/1126-6708/2004/04/048;%%
  %209 citations counted in INSPIRE as of 16 Dec 2016

%\cite{Cariglia:2004kk}
\bibitem{Cariglia:2004kk}
  M.~Cariglia and O.~A.~P.~Mac Conamhna,
  ``The General form of supersymmetric solutions of N=(1,0) U(1) and SU(2) gauged supergravities in six-dimensions,''
  Class.\ Quant.\ Grav.\  {\bf 21} (2004) 3171
  %doi:10.1088/0264-9381/21/13/006
  [hep-th/0402055].
  %%CITATION = doi:10.1088/0264-9381/21/13/006;%%
  %65 citations counted in INSPIRE as of 17 Nov 2016

%\cite{Cacciatori:2004rt}
\bibitem{Cacciatori:2004rt}
  S.~L.~Cacciatori, M.~M.~Caldarelli, D.~Klemm and D.~S.~Mansi,
  ``More on BPS solutions of N = 2, D = 4 gauged supergravity,''
  JHEP {\bf 0407} (2004) 061
  %doi:10.1088/1126-6708/2004/07/061
  [hep-th/0406238].
  %%CITATION = doi:10.1088/1126-6708/2004/07/061;%%
  %48 citations counted in INSPIRE as of 17 Nov 2016

%\cite{Bellorin:2007yp}
\bibitem{Bellorin:2007yp}
  J.~Bellorin and T.~Ortin,
  ``Characterization of all the supersymmetric solutions of gauged N=1, d=5 supergravity,''
  JHEP {\bf 0708} (2007) 096
  %doi:10.1088/1126-6708/2007/08/096
  [arXiv:0705.2567 [hep-th]].
  %%CITATION = doi:10.1088/1126-6708/2007/08/096;%%
  %35 citations counted in INSPIRE as of 17 Nov 2016

%\cite{Cacciatori:2008ek}
\bibitem{Cacciatori:2008ek}
  S.~L.~Cacciatori, D.~Klemm, D.~S.~Mansi and E.~Zorzan,
  ``All timelike supersymmetric solutions of N=2, D=4 gauged supergravity coupled to abelian vector multiplets,''
  JHEP {\bf 0805} (2008) 097
  %doi:10.1088/1126-6708/2008/05/097
  [arXiv:0804.0009 [hep-th]].
  %%CITATION = doi:10.1088/1126-6708/2008/05/097;%%
  %49 citations counted in INSPIRE as of 17 Nov 2016

%\cite{Hubscher:2008yz}
\bibitem{Hubscher:2008yz}
  M.~Huebscher, P.~Meessen, T.~Ortin and S.~Vaula,
  ``N=2 Einstein-Yang-Mills's BPS solutions,''
  JHEP {\bf 0809} (2008) 099
  %doi:10.1088/1126-6708/2008/09/099
  [arXiv:0806.1477 [hep-th]].
  %%CITATION = doi:10.1088/1126-6708/2008/09/099;%%
  %45 citations counted in INSPIRE as of 17 Nov 2016

%\cite{Bellorin:2008we}
\bibitem{Bellorin:2008we}
  J.~Bellorin,
  ``Supersymmetric solutions of gauged five-dimensional supergravity with general matter couplings,''
  Class.\ Quant.\ Grav.\  {\bf 26} (2009) 195012
   [arXiv:0810.0527 [hep-th]].
  %%CITATION = doi:10.1088/0264-9381/26/19/195012;%%
  %12 citations counted in INSPIRE as of 23 Jun 2016

%\cite{Klemm:2009uw}
\bibitem{Klemm:2009uw}
  D.~Klemm and E.~Zorzan,
  ``All null supersymmetric backgrounds of N=2, D=4 gauged supergravity coupled to abelian vector multiplets,''
  Class.\ Quant.\ Grav.\  {\bf 26} (2009) 145018
  %doi:10.1088/0264-9381/26/14/145018
  [arXiv:0902.4186 [hep-th]].
  %%CITATION = doi:10.1088/0264-9381/26/14/145018;%%
  %29 citations counted in INSPIRE as of 17 Nov 2016

%\cite{Hristov:2009uj}
\bibitem{Hristov:2009uj}
  K.~Hristov, H.~Looyestijn and S.~Vandoren,
  ``Maximally supersymmetric solutions of D=4 N=2 gauged supergravity,''
  JHEP {\bf 0911} (2009) 115
  [arXiv:0909.1743 [hep-th]].
  %%CITATION = doi:10.1088/1126-6708/2009/11/115;%%
  %35 citations counted in INSPIRE as of 23 Jun 2016

%\cite{Klemm:2010mc}
\bibitem{Klemm:2010mc}
  D.~Klemm and E.~Zorzan,
  ``The timelike half-supersymmetric backgrounds of N=2, D=4 supergravity with Fayet-Iliopoulos gauging,''
  Phys.\ Rev.\ D {\bf 82} (2010) 045012
  %doi:10.1103/PhysRevD.82.045012
  [arXiv:1003.2974 [hep-th]].
  %%CITATION = doi:10.1103/PhysRevD.82.045012;%%
  %17 citations counted in INSPIRE as of 17 Nov 2016

%\cite{Meessen:2012sr}
\bibitem{Meessen:2012sr}
  P.~Meessen and T.~Ortin,
  ``Supersymmetric solutions to gauged N=2 d=4 sugra: the full timelike shebang,''
  Nucl.\ Phys.\ B {\bf 863} (2012) 65
  %doi:10.1016/j.nuclphysb.2012.05.023
  [arXiv:1204.0493 [hep-th]].
  %%CITATION = doi:10.1016/j.nuclphysb.2012.05.023;%%
  %26 citations counted in INSPIRE as of 17 Nov 2016














\bibitem{Freund:1980xh} 
  P.~G.~O.~Freund and M.~A.~Rubin,
  ``Dynamics of Dimensional Reduction,''
  Phys.\ Lett.\ B {\bf 97}, 233 (1980).
 % doi:10.1016/0370-2693(80)90590-0
  %%CITATION = doi:10.1016/0370-2693(80)90590-0;%%
  %768 citations counted in INSPIRE as of 19 Feb 2016





%\cite{KowalskiGlikman:1984wv}
\bibitem{KowalskiGlikman:1984wv}
  J.~Kowalski-Glikman,
  ``Vacuum States in Supersymmetric Kaluza-Klein Theory,''
  Phys.\ Lett.\ B {\bf 134} (1984) 194.
  %%CITATION = doi:10.1016/0370-2693(84)90669-5;%%
  %153 citations counted in INSPIRE as of 14 Jul 2016

%\cite{KowalskiGlikman:1985im}
\bibitem{KowalskiGlikman:1985im}
  J.~Kowalski-Glikman,
  ``Positive Energy Theorem And Vacuum States For The Einstein-maxwell System,''
  Phys.\ Lett.\ B {\bf 150} (1985) 125.
  %%CITATION = doi:10.1016/0370-2693(85)90153-4;%%
  %31 citations counted in INSPIRE as of 14 Jul 2016




%\cite{Samtleben:2008pe}
\bibitem{Samtleben:2008pe}
  H.~Samtleben,
  ``Lectures on Gauged Supergravity and Flux Compactifications,''
  Class.\ Quant.\ Grav.\  {\bf 25} (2008) 214002
%  doi:10.1088/0264-9381/25/21/214002
  [arXiv:0808.4076 [hep-th]].
  %%CITATION = doi:10.1088/0264-9381/25/21/214002;%%
  %106 citations counted in INSPIRE as of 12 Apr 2016

%\cite{Bandos:2016smv}
\bibitem{Bandos:2016smv}
  I.~A.~Bandos and T.~Ortin,
  ``On the dualization of scalars into (d - 2)-forms in supergravity. Momentum maps, R-symmetry and gauged supergravity,''
  JHEP {\bf 1608} (2016) 135
  %doi:10.1007/JHEP08(2016)135
  [arXiv:1605.05559 [hep-th]].
  %%CITATION = doi:10.1007/JHEP08(2016)135;%%
  %3 citations counted in INSPIRE as of 15 Nov 2016

%\cite{Trigiante:2016mnt}
\bibitem{Trigiante:2016mnt}
  M.~Trigiante,
  ``Gauged Supergravities,''
  arXiv:1609.09745 [hep-th].
  %%CITATION = ARXIV:1609.09745;%%

%\cite{deAlwis:2013jaa}
\bibitem{deAlwis:2013jaa}
  S.~de Alwis, J.~Louis, L.~McAllister, H.~Triendl and A.~Westphal,
  ``Moduli spaces in $AdS_4$ supergravity,''
  JHEP {\bf 1405} (2014) 102
  [arXiv:1312.5659 [hep-th]].
  %%CITATION = ARXIV:1312.5659;%%
  %1 citations counted in INSPIRE as of 05 Aug 2014


%\cite{Louis:2014gxa}
\bibitem{Louis:2014gxa}
  J.~Louis and H.~Triendl,
  ``Maximally supersymmetric AdS$_{4}$ vacua in N = 4 supergravity,''
  JHEP {\bf 1410} (2014) 007
  [arXiv:1406.3363 [hep-th]].
  %%CITATION = ARXIV:1406.3363;%%


%\cite{Louis:2015mka}
\bibitem{Louis:2015mka}
  J.~Louis and S.~L\"ust,
  ``Supersymmetric AdS$_{7}$ backgrounds in half-maximal supergravity and marginal operators of (1, 0) SCFTs,''
  JHEP {\bf 1510} (2015) 120
%  doi:10.1007/JHEP10(2015)120
  [arXiv:1506.08040 [hep-th]].
  %%CITATION = doi:10.1007/JHEP10(2015)120;%%
  %4 citations counted in INSPIRE as of 20 Feb 2016




%\cite{Louis:2015dca}
\bibitem{Louis:2015dca}
  J.~Louis, H.~Triendl and M.~Zagermann,
  ``$ \mathcal{N}=4 $ supersymmetric AdS$_{5}$ vacua and their moduli spaces,''
  JHEP {\bf 1510} (2015) 083
%  doi:10.1007/JHEP10(2015)083
  [arXiv:1507.01623 [hep-th]].
  %%CITATION = doi:10.1007/JHEP10(2015)083;%%
  %4 citations counted in INSPIRE as of 20 Feb 2016

%\cite{Louis:2016qca}
\bibitem{Louis:2016qca}
  J.~Louis and C.~Muranaka,
  ``Moduli spaces of AdS$_{5}$ vacua in $ \mathcal{N} $ = 2 supergravity,''
  JHEP {\bf 1604} (2016) 178
  [arXiv:1601.00482 [hep-th]].
  %%CITATION = doi:10.1007/JHEP04(2016)178;%%
  %2 citations counted in INSPIRE as of 25 Jul 2016


%\cite{Nahm:1977tg}
\bibitem{Nahm:1977tg}
  W.~Nahm,
  ``Supersymmetries and their Representations,''
  Nucl.\ Phys.\ B {\bf 135} (1978) 149.
  %%CITATION = doi:10.1016/0550-3213(78)90218-3;%%
  %553 citations counted in INSPIRE as of 15 Apr 2016


%\cite{Schwarz:1983qr}
\bibitem{Schwarz:1983qr}
  J.~H.~Schwarz,
  ``Covariant Field Equations of Chiral N=2 D=10 Supergravity,''
  Nucl.\ Phys.\ B {\bf 226} (1983) 269.
 % doi:10.1016/0550-3213(83)90192-X
  %%CITATION = doi:10.1016/0550-3213(83)90192-X;%%
  %653 citations counted in INSPIRE as of 22 Mar 2016


%\cite{Schwarz:1983wa}
\bibitem{Schwarz:1983wa}
  J.~H.~Schwarz and P.~C.~West,
  ``Symmetries and Transformations of Chiral N=2 D=10 Supergravity,''
  Phys.\ Lett.\ B {\bf 126} (1983) 301.
  %%CITATION = doi:10.1016/0370-2693(83)90168-5;%%
  %300 citations counted in INSPIRE as of 25 Jul 2016




%\cite{Nishino:1984gk}
\bibitem{Nishino:1984gk}
  H.~Nishino and E.~Sezgin,
  ``Matter and Gauge Couplings of N=2 Supergravity in Six-Dimensions,''
  Phys.\ Lett.\ B {\bf 144} (1984) 187.
 % doi:10.1016/0370-2693(84)91800-8
  %%CITATION = doi:10.1016/0370-2693(84)91800-8;%%
  %191 citations counted in INSPIRE as of 22 Mar 2016

%\cite{Awada:1985er}
\bibitem{Awada:1985er}
  M.~Awada, P.~K.~Townsend and G.~Sierra,
  ``Six-dimensional Simple and Extended Chiral Supergravity in Superspace,''
  Class.\ Quant.\ Grav.\  {\bf 2} (1985) L85.
  %%CITATION = doi:10.1088/0264-9381/2/4/005;%%
  %15 citations counted in INSPIRE as of 25 Jul 2016







\bibitem{Cahen1970}
  M.~Cahen and N.~Wallach,
  ``Lorentzian symmetric spaces,''
  Bull.\ Am.\ Math.\ Soc.\ {\bf 76} (1970) 585-591.



\bibitem{Penrose1976}
  R.~Penrose,
  ``Any space-time has a plane wave as a limit,"
  in \textit{Differential geometry and relativity}, Reidel, Dordrecht (1976) 271-275.

%\cite{Gueven:2000ru}
\bibitem{Gueven:2000ru}
  R.~Gueven,
  ``Plane wave limits and T duality,''
  Phys.\ Lett.\ B {\bf 482} (2000) 255
  [hep-th/0005061].
  %%CITATION = doi:10.1016/S0370-2693(00)00517-7;%%
  %206 citations counted in INSPIRE as of 14 Jul 2016

%\cite{Blau:2002dy}
\bibitem{Blau:2002dy}
  M.~Blau, J.~M.~Figueroa-O'Farrill, C.~Hull and G.~Papadopoulos,
  ``Penrose limits and maximal supersymmetry,''
  Class.\ Quant.\ Grav.\  {\bf 19} (2002) L87
  [hep-th/0201081].
  %%CITATION = doi:10.1088/0264-9381/19/10/101;%%
  %467 citations counted in INSPIRE as of 14 Jul 2016

%\cite{Blau:2002mw}
\bibitem{Blau:2002mw}
  M.~Blau, J.~M.~Figueroa-O'Farrill and G.~Papadopoulos,
  ``Penrose limits, supergravity and brane dynamics,''
  Class.\ Quant.\ Grav.\  {\bf 19} (2002) 4753
   [hep-th/0202111].
  %%CITATION = doi:10.1088/0264-9381/19/18/310;%%
  %255 citations counted in INSPIRE as of 14 Jul 2016

%\cite{Breckenridge:1996is}
\bibitem{Breckenridge:1996is}
  J.~C.~Breckenridge, R.~C.~Myers, A.~W.~Peet and C.~Vafa,
  ``D-branes and spinning black holes,''
  Phys.\ Lett.\ B {\bf 391} (1997) 93
   [hep-th/9602065].
  %%CITATION = doi:10.1016/S0370-2693(96)01460-8;%%
  %462 citations counted in INSPIRE as of 13 Jul 2016

%\cite{Fiol:2003yq}
\bibitem{Fiol:2003yq}
  B.~Fiol, C.~Hofman and E.~Lozano-Tellechea,
  ``Causal structure of d = 5 vacua and axisymmetric space-times,''
  JHEP {\bf 0402} (2004) 034
   [hep-th/0312209].
  %%CITATION = doi:10.1088/1126-6708/2004/02/034;%%
  %9 citations counted in INSPIRE as of 13 Jul 2016


%\cite{AlonsoAlberca:2002wr}
\bibitem{AlonsoAlberca:2002wr}
  N.~Alonso-Alberca, E.~Lozano-Tellechea and T.~Ortin,
  %``The Near horizon limit of the extreme rotating D = 5 black hole as a homogeneous space-time,''
  Class.\ Quant.\ Grav.\  {\bf 20} (2003) 423
  [hep-th/0209069].
  %%CITATION = doi:10.1088/0264-9381/20/3/303;%%
  %17 citations counted in INSPIRE as of 14 Jul 2016

%\cite{LozanoTellechea:2002pn}
\bibitem{LozanoTellechea:2002pn}
  E.~Lozano-Tellechea, P.~Meessen and T.~Ortin,
  ``On d = 4, d = 5, d = 6 vacua with eight supercharges,''
  Class.\ Quant.\ Grav.\  {\bf 19} (2002) 5921
  [hep-th/0206200].
  %%CITATION = doi:10.1088/0264-9381/19/23/303;%%
  %30 citations counted in INSPIRE as of 14 Jul 2016



% %\cite{Biran:1982eg}
% \bibitem{Biran:1982eg}
%   B.~Biran, F.~Englert, B.~de Wit and H.~Nicolai,
%   ``Gauged $N=8$ Supergravity and Its Breaking From Spontaneous Compactification,''
%   Phys.\ Lett.\ B {\bf 124} (1983) 45
%    Erratum: [Phys.\ Lett.\ B {\bf 128} (1983) 461].
% %  doi:10.1016/0370-2693(83)91400-4
%   %%CITATION = doi:10.1016/0370-2693(83)91400-4;%%
%   %115 citations counted in INSPIRE as of 15 Apr 2016



%\cite{Blau:2001ne}
\bibitem{Blau:2001ne}
  M.~Blau, J.~M.~Figueroa-O'Farrill, C.~Hull and G.~Papadopoulos,
  ``A New maximally supersymmetric background of IIB superstring theory,''
  JHEP {\bf 0201} (2002) 047
  [hep-th/0110242].
  %%CITATION = doi:10.1088/1126-6708/2002/01/047;%%
  %531 citations counted in INSPIRE as of 26 Jul 2016

%\cite{Gibbons:1994vm}
\bibitem{Gibbons:1994vm}
  G.~W.~Gibbons, G.~T.~Horowitz and P.~K.~Townsend,
  ``Higher dimensional resolution of dilatonic black hole singularities,''
  Class.\ Quant.\ Grav.\  {\bf 12} (1995) 297
  [hep-th/9410073].
  %%CITATION = doi:10.1088/0264-9381/12/2/004;%%
  %273 citations counted in INSPIRE as of 26 Jul 2016

%\cite{Meessen:2001vx}
\bibitem{Meessen:2001vx}
  P.~Meessen,
  ``A Small note on P P wave vacua in six-dimensions and five-dimensions,''
  Phys.\ Rev.\ D {\bf 65} (2002) 087501
  [hep-th/0111031].
  %%CITATION = doi:10.1103/PhysRevD.65.087501;%%
  %44 citations counted in INSPIRE as of 26 Jul 2016

%\cite{Chamseddine:1996pi}
\bibitem{Chamseddine:1996pi}
  A.~H.~Chamseddine, S.~Ferrara, G.~W.~Gibbons and R.~Kallosh,
  ``Enhancement of supersymmetry near 5-d black hole horizon,''
  Phys.\ Rev.\ D {\bf 55} (1997) 3647
  [hep-th/9610155].
  %%CITATION = doi:10.1103/PhysRevD.55.3647;%%
  %95 citations counted in INSPIRE as of 26 Jul 2016

%\cite{Cvetic:1998xh}
\bibitem{Cvetic:1998xh}
  M.~Cvetic and F.~Larsen,
  ``Near horizon geometry of rotating black holes in five-dimensions,''
  Nucl.\ Phys.\ B {\bf 531} (1998) 239
  [hep-th/9805097].
  %%CITATION = doi:10.1016/S0550-3213(98)00604-X;%%
  %112 citations counted in INSPIRE as of 26 Jul 2016

%\cite{Gauntlett:1998fz}
\bibitem{Gauntlett:1998fz}
  J.~P.~Gauntlett, R.~C.~Myers and P.~K.~Townsend,
  ``Black holes of D = 5 supergravity,''
  Class.\ Quant.\ Grav.\  {\bf 16} (1999) 1
  [hep-th/9810204].
  %%CITATION = doi:10.1088/0264-9381/16/1/001;%%
  %237 citations counted in INSPIRE as of 26 Jul 2016

%\cite{Bertotti:1959pf}
\bibitem{Bertotti:1959pf}
  B.~Bertotti,
  ``Uniform electromagnetic field in the theory of general relativity,''
  Phys.\ Rev.\  {\bf 116} (1959) 1331.
   %%CITATION = doi:10.1103/PhysRev.116.1331;%%
  %255 citations counted in INSPIRE as of 26 Jul 2016

%\cite{Robinson:1959ev}
\bibitem{Robinson:1959ev}
  I.~Robinson,
  ``A Solution of the Maxwell-Einstein Equations,''
  Bull.\ Acad.\ Pol.\ Sci.\ Ser.\ Sci.\ Math.\ Astron.\ Phys.\  {\bf 7} (1959) 351.
  %%CITATION = BAPMA,7,351;%%
  %125 citations counted in INSPIRE as of 26 Jul 2016

%\cite{AlonsoAlberca:2002dw}
\bibitem{AlonsoAlberca:2002dw}
  N.~Alonso-Alberca and T.~Ortin,
  ``Supergravity vacua today,''
  gr-qc/0210039.
  %%CITATION = GR-QC/0210039;%%
  %2 citations counted in INSPIRE as of 26 Jul 2016

%\cite{Freedman:2012zz}
\bibitem{Freedman:2012zz} 
  D.~Z.~Freedman and A.~Van Proeyen,
  ``Supergravity,'' Cambridge University Press, 2012.
  %%CITATION = INSPIRE-1123253;%%
  %22 citations counted in INSPIRE as of 21 Mar 2016


% %\cite{Pilch:1984xy}
% \bibitem{Pilch:1984xy}
%   K.~Pilch, P.~van Nieuwenhuizen and P.~K.~Townsend,
%   ``Compactification of $d=11$ Supergravity on S(4) (Or 11 = 7 + 4, Too),''
%   Nucl.\ Phys.\ B {\bf 242} (1984) 377.
%  % doi:10.1016/0550-3213(84)90400-0
%   %%CITATION = doi:10.1016/0550-3213(84)90400-0;%%
%   %122 citations counted in INSPIRE as of 15 Apr 2016




% %\cite{Kim:1985ez}
% \bibitem{Kim:1985ez}
%   H.~J.~Kim, L.~J.~Romans and P.~van Nieuwenhuizen,
%   ``The Mass Spectrum of Chiral N=2 D=10 Supergravity on S**5,''
%   Phys.\ Rev.\ D {\bf 32} (1985) 389.
%  % doi:10.1103/PhysRevD.32.389
%   %%CITATION = doi:10.1103/PhysRevD.32.389;%%
%   %455 citations counted in INSPIRE as of 15 Apr 2016


\end{thebibliography}
\end{document}